\documentclass[fleqn,usenatbib]{mnras}
\usepackage{newtxtext,newtxmath}

\usepackage[T1]{fontenc}
\usepackage{ae,aecompl}
\usepackage{lscape}

\usepackage{graphicx}	% Including figure files
\usepackage{amsmath}	% Advanced maths commands
\usepackage[caption=false]{subfig}

\newcommand{\starname}{HD\,18599}

\newcommand{\tess}{\textit{TESS}}

\newcommand{\species}{\texttt{SPECIES}}
\newcommand{\ariadne}{\texttt{ARIADNE}}

\newcommand{\exofast}{\texttt{EXOFASTv2}}

\newcommand{\me}{$M_{\oplus}$}
\newcommand{\re}{$R_{\oplus}$}

\newcommand{\kms}{km\,s$^{-1}$}
\newcommand{\ms}{m\,s$^{-1}$}

\newcommand{\rstar}{\ensuremath{{\rm R}_{\star}}}
\newcommand{\mstar}{\ensuremath{{\rm M}_{\star}}}

\newcommand{\mpl}{\ensuremath{{\rm M_p}}}
\newcommand{\rpl}{\ensuremath{{\rm R_p}}}

\newcommand{\masy}{mas\,y$^{-1}$}
\newcommand{\msun}{\mbox{$M_{\odot}$}}
\newcommand{\rsun}{\mbox{$R_{\odot}$}}

\newcommand{\gccc}{g\,cm$^{-3}$}

\title[\starname b]{A dense mini-Neptune orbiting the bright young star \starname}

\author[Vines, J. I. et al.]{
\parbox{\textwidth}{
Jose I. Vines,$^{1}$\thanks{E-mail: \href{jose.vines@ug.uchile.cl}{jose.vines@ug.uchile.cl}}
James S. Jenkins,$^{2,3}$
Zaira Berdi\~nas,$^{1}$
Maritza G.~Soto,$^{4}$
Mat{\'i}as R. D{\'i}az,$^{1,5}$
Douglas R. Alves,$^{1}$
Mikko Tuomi,$^{9}$
Robert A. Wittenmyer,$^{6}$  
Jerome Pitogo de Leon,$^{7}$
Pablo Pe\~na,$^{2}$
Jack J. Lissauer,$^{8}$
Sarah Ballard,$^{10}$
Timothy Bedding,$^{11}$
Brendan P. Bowler,$^{12}$
Jonathan Horner,$^{6}$  
Hugh R.A. Jones,$^{13}$
Stephen R. Kane,$^{14}$  
John Kielkopf,$^{15}$
Peter Plavchan,$^{16}$
Avi Shporer,$^{17}$
C. G. Tinney,$^{18}$
Hui Zhang$^{19}$
Duncan J. Wright,$^{6}$   %0000-0001-7294-5386
Brett Addison,$^{6}$
Matthew W. Mengel,$^{6}$
Jack Okumura,$^{6}$
Anya Samadi-Ghadim,$^{20}$
}
\\
$^{1}$Departamento de Astronom\'ia, Universidad de Chile, Casilla 36-D, Santiago, Chile\\
$^{2}$ N\'ucleo de Astronom\'ia, Facultad de Ingenier\'ia y Ciencias, Universidad Diego Portales, Av. Ej\'ercito 441, Santiago, Chile\\
$^{3}$Centro de Astrof\'isica y Tecnolog\'ias Afines (CATA), Casilla 36-D, Santiago, Chile\\
$^{4}$School of Physics and Astronomy, Queen Mary University of London, 327 Mile End Road, London E1 4NS, UK\\
$^{5}$ Las Campanas Observatory, Carnegie Institution of Washington, Colina El Pino, Casilla 601 La Serena, Chile.\\
$^{6}$University of Southern Queensland, Centre for Astrophysics, West Street, Toowoomba, QLD 4350 Australia\\
$^{7}$Department of Astronomy, The University of Tokyo, 7-3-1 Hongo, Bunkyo-ku, Tokyo 113-0033, Japan\\
$^{8}$NASA Ames Research Center, Moffett Field, CA, 94035, USA\\
$^{9}$Department of Physics, University of Helsinki, PO Box 64, 00014, Finland\\
$^{10}$Department of Astronomy, University of Florida, 211 Bryant Space Science Center, Gainesville, FL, 32611, USA\\
$^{11}$School of Physics, Sydney Institute for Astronomy (SIfA), The University of Sydney, NSW 2006, Australia\\
$^{12}$Department of Astronomy, The University of Texas at Austin, TX 78712, USA\\
$^{13}$School of Physics, Astronomy and Mathematics, University of Hertfordshire, College Lane, Hatfield, AL10 9AB, UK\\
$^{14}$Department of Earth and Planetary Sciences, University of California, Riverside, CA 92521, USA\\
$^{15}$Department of Physics and Astronomy, University of Louisville, Louisville, KY 40292, USA\\
$^{16}$George Mason University, 4400 University Drive MS 3F3, Fairfax, VA 22030, USA\\
$^{17}$Department of Physics and Kavli Institute for Astrophysics and Space Research, Massachusetts Institute of Technology, Cambridge, MA 02139, USA\\
$^{18}$Exoplanetary Science at UNSW, School of Physics, UNSW Sydney, NSW 2052, Australia\\
$^{19}$School of Astronomy and Space Science, Key Laboratory of Modern Astronomy and Astrophysics in Ministry of Education, Nanjing University, Nanjing 210046, Jiangsu, China\\
$^{20}$Max-Planck-Institut für Sonnensystemforschung, 37077 Göttingen, Germany}

\date{in prep.}

\pubyear{20NN}

\begin{document}
\label{firstpage}
\pagerange{\pageref{firstpage}--\pageref{lastpage}}
\maketitle

\begin{abstract}

Very little is known about the young planet population because the detection of small planets orbiting young stars is obscured by the effects of stellar activity and fast rotation which mask planets within radial velocity and transit data sets. The few planets that have been discovered in young clusters generally orbit stars too faint for any detailed follow-up analysis. Here we present the characterization of a new mini-Neptune planet orbiting the bright (V=9) and nearby K2 dwarf star, \starname.  The planet candidate was originally detected in \tess{}\, light curves from Sectors 2, 3, 29, and 30, with an orbital period of 4.138~days. We then used HARPS and FEROS radial velocities, to find the companion mass to be 25.5$\pm$4.6~\me. When we combine this with the measured radius from \tess{}\, of 2.70$\pm$0.05~\re, we find a high planetary density of 7.1$\pm$1.4~\gccc.  The planet exists on the edge of the Neptune Desert and is the first young planet (300 Myr) of its type to inhabit this region.  Structure models argue for a bulk composition to consist of 23\% H$_2$O and 77\% Rock and Iron.  Future follow-up with large ground- and space-based telescopes can enable us to begin to understand in detail the characteristics of young Neptunes in the galaxy.

\end{abstract}


%%%%%%%%%%%%%%%%%%%%%%%%%%%%%%%%%%%%%%%%%%%%%%%%%%

%%%%%%%%%%%%%%%%% BODY OF PAPER %%%%%%%%%%%%%%%%%%

\section{Introduction}
\label{sec:intro}

The Neptune Desert, also known as the sub-Jovian Desert, is a region in the period-mass-radius planetary parameter space relatively devoid of planets \citep{neptune_desert}.  The trapezoidal shape of the desert was outlined by \citet{neptune-desert-edges}, with two firm edges that were easily identifiable in results drawn from Kepler Space Telescope \citep{KEPLER}.  It seems that planets closest to their host stars come in two flavours, either large gas giants, or small rocky planets.  The desert can therefore be explained by the combined effects of tidal migration and planetary photoevaporation by the host star \citep{Owen_Wu_18}.  Giant planets are too massive to be significantly affected by the high-energy XUV radiation from the host star, no matter how close to their host star they migrate, and therefore they maintain a dominant gaseous envelope.  Planets like Neptune, on the other hand, can not hold onto their gaseous envelopes, losing them early in their evolution, and ending up as left-over hot cores, with minimal residual atmospheres.  This mechanism naturally explains the Neptune Desert region.

Although the Neptune Desert is 'dry', it is not completely devoid of planets.  Recently a few planets have been discovered residing close to their stars, yet with masses and radii in the Neptune regime.  One of the first planets announced from the Next Generation Planet Search \citep[NGTS,][]{NGTS}, NGTS-4b \citep{NGTS-4b}, was a dense planet found to have a period, mass, and radius that placed it in the Neptune Desert.  Subsequently, \tess{} detected the planet TOI-849b \citep{TOI-849}, another dense super-Neptune within the desert.  This planet likely hosts no atmosphere at all, probably the result of a previous gas giant that was entirely stripped of its gaseous envelope. A further \tess{} discovery was planet orbiting LTT\,9779 a unique desert Neptune, having an orbital period of 19~hrs, it maintains an equilibrium temperature above 2000~K, making it the first Ultrahot Neptune \citep{LTT9779}.  More intruiging is the fact that this planet does host a significant atmosphere \citep[see][]{dragomir20, crossfield20}, even though the system is not young. Given that the Neptune desert is well defined empirically and explained by models, the study of objects within the desert is particularly intriguing though bright examples to test photoevaporation models are needed.

Since photoevaporation appears to be the mechanism that gives rise to the desert, we expect the majority of mass-loss to happen very early in the evolution of the system.  Stars emit the majority of their high-energy X-ray and UV radiation in the first 100~Myrs of their life \citep{Jackson2012}, the photons that heat the upper atmospheres of planets sufficiently to cause gas to escape the planet's gravitational potential.  Other models that can explain planets to lose significant amounts of their atmospheres, generally happen on longer timescales, like core driven mass-loss \citep{GuptaSchichting20} or Roche Lobe Overflow \citep{Valsecchi2014}.  Therefore, studying this region as a function of age will not only shed light on the dominant processes, but also allow a more detailed characterisation of the timescales these processes operate over.

\subsection{Significance of \starname}

The southern star \starname\, (TOI-179) has a $V$ magnitude of 8.99 mags, and had an estimated spectral type of K2V based on data from the Hipparcos mission \citep{HIPPARCOS, NEW_HIP_REDUCTION}.  The Gaia EDR3 parallax gives rise to a distance of 38.632~pc \citep{GAIA_EDR3}, and given the brightness, this is consistent with the aboslute magnitude expected for a K-dwarf.  The star was screened as part of the first phases of the Calan-Hertfordshire Extrasolar Planet Search project (CHEPS; \citealp{Jenkins2009, Jenkins2017}) through measurements of the stellar chromospheric activity, metallicity, rotational velocity, and kinematics.  They found the star to have a log$R'_{\rm{HK}}$ index of -4.39~dex \citep{Jenkins2008, Jenkins2011}, $v$sin($i$) of 4.3~\kms, and kinematic U, V, and W velocities of -3.49, 8.50, and -6.21~\kms, respectively \citep{Murgas2013}.  Applying gyrochronological-based relations to these measured values (e.g. see \citealt{Mamajek2008}), provides estimates of the age of the system, coming in at around 300~Myrs.

Although the star was recognised to be too active for inclusion in the CHEPS precision radial velocity program, (since that study focused only on the most chromospherically quiet stars), the Transiting Exoplanet Survey Satellite \citep[\tess{},][]{TESS} mission does not have the same activity constraints, and a transiting planet candidate was detected as part of that mission.  The candidate was subsequently validated as a planet by de Leon et al. 2022, in prep., providing another possible planet on the edge of the Neptune Desert.  However, unlike most of the other systems in this region, we can confidently say this is younger than typical field stars, and given the brightness of the host, it could allow more stringent constraints to be placed on the timescales of the photoevaporative process that is believed to drive the existence of the desert, along with allowing atmospheric characterisation of the younger cohort of planets.

%%%%%%%%%%%%%%%%%%%%%%%%%%%%%%%%%%%%%%%%%%%%%%%%%%%%%%%%%%%%%%%%%%%%%%%%%%%%%%%%%%%%%%%%%%%%%%%%%%%%%%%%%%%%%%%%%%%%%%%%%%%%%%%%%%%%%%%%%%%%%%%%%%%%%%%%%%%%%%%%%%

\section{Observations}
\label{sec:obs}

This candidate was first detected through the observation of 16 transit events by the \tess{} mission, yielding a period of 4.137~days from an \exofast\, \citep{exofastv2} analysis listed on the Exoplanet Follow-up Observing Program (ExoFOP)\footnote{https://exofop.ipac.caltech.edu/tess/target.php?id=207141131}. After scrutinizing the processed light curve publicly available at MAST\footnote{https://mast.stsci.edu/portal/Mashup/Clients/Mast/Portal.html}, and given that the star is bright, we selected this candidate as a high-priority target to be observed in a HARPS campaign starting in 2018. Upon examination of archival data we found the star was already observed with sparse sampling between 2014 and 2017. Later, additional observations with FEROS were planned to help constrain the orbital and physical parameters of the planet. Finally, Minerva-Australis observations to help validate the candidate were also acquired. The following sections describe the observations and data acquisition of HD 18599 in detail.
%%%%%%%%%%%%%%%%%%%%%%%%%%%%%%%%%%%%%%%%%%%%%%%%%%%%%%%%%%%%%%%%%%%%%%%%%%%%%%%%%

\subsection{Photometry}

\subsubsection{WASP Photometry}
\label{sec:waspphot}

WASP-South, an array of 8 wide-field cameras, was the Southern station of the WASP transit-search project \citep{2006PASP..118.1407P}. It observed the field of \starname\, in 2010 and 2011, when equipped with 200-mm, f/1.8 lenses, and then in 2012, 2013 and 2014, equipped with 85-mm, f/1.2 lenses. Observations were made on each clear night over an observing season of 175 nights per year, accumulating 86\,000 photometric data points with a typical 15-min cadence. At $V$ = 9, \starname\, is by far the brightest star in the 48-arcsec extraction aperture of the 200-mm data.  A $V$ = 10.9 star, 88 arcsecs away, is, however, within the 113-arcsec extraction aperture of the 85-mm data, and will cause a $\sim$\,10\%\ dilution. We studied the periodogram of the WASP light curves in Section \ref{sub:rotation} to find the rotation period of \starname\, in order to properly detrend the \tess{} data.

\subsubsection{\tess{} Photometry}
\label{sec:tessphot}

HD 18599 was observed by \tess{} in Sectors 2, 3, 29, and 30 with Camera 3 in short-cadence mode (2 minutes), starting from the 22nd of August 2018 to the 22nd of September 2020, producing a total time baseline of 2 years, during which 21 transit features were observed. Observations were analyzed and processed with the Science Processing Operations Center \citep[SPOC,][]{spoc} by NASA Ames Research Center, after which it was promoted to TOI-179.

We extracted the Simple Aperture Photometry (SAP) and the Presearch Data Conditioning \citep[PDC,][]{spoc} light curves that were produced by the SPOC pipeline, removed data points flagged as low quality, and finally we normalized the light curve. We note that even after this procedure, stellar variability still remains in the light curve and must be taken care of to determine the most optimal companion parameters. In the end, the SAP and PDCSAP light curves look almost identical, but we use the latter throughout the paper in any case.

Finally we applied Gaussian Processes (GP) to detrend and remove the stellar variability signal in the light curve (see Section \ref{sec:activity} and Figure \ref{fig:detrend}). The detrended light curve can be found in Table \ref{tab:tesslc}.

\begin{table}
	\centering
	\caption{Detrended \tess{} photometry for HD 18599.  The full table is available in a machine-readable format from the online journal.  A portion is shown here for guidance.}
	\label{tab:tesslc}
	\begin{tabular}{ccc}
	Time	&	Flux        	&Flux\\
    (BJD-2450000)	&	(normalised)	&error\\
	\hline
    8354.11157  &  0.9989  &  0.0004 \\
	8354.11296  &  0.9992  &  0.0004 \\
	8354.11574  &  0.9993  &  0.0004 \\
	8354.11712  &  0.9994  &  0.0004 \\
	8354.11851  &  0.9994  &  0.0004 \\
	8354.11990  &  0.9999  &  0.0004 \\
	8354.12129  &  0.9990  &  0.0004 \\
	8354.12268  &  0.9993  &  0.0004 \\
	8354.12407  &  0.9993  &  0.0004 \\
	8354.12546  &  0.9995  &  0.0004 \\
    \vdots      &  \vdots  &  \vdots \\
	\hline
	\end{tabular}
\end{table}

%%%%%%%%%%%%%%%%%%%%%%%%%%%%%%%%%%%%%%%%%%%%%%%%%%%%%%%%%%%%%%%%%%%%%%%%%%%%%%%%%

\subsection{Radial Velocities}
\label{sub:spect}
We observed HD 18599 with three fiber-fed high precision \'echelle spectrographs: High-Accuray Radial velocity Planet Searcher \cite[HARPS,][]{Pepe2002}, FEROS, and Minerva-Australis to obtain high-precision radial velocities (RVs) and constrain the planetary mass.

\subsubsection{HARPS}
HARPS is mounted on the ESO 3.6-m Telescope at the La Silla observatory in Northern Chile, and has a spectral resolution of R = 120 000. Observations include data acquired prior to the fibre updates we refer to as "pre" (PI: Lagrange). Data acquired after the fiber updates in 2015 (PI: Lagrange, D\'iaz, Berdi\~{n}as, Jord\'an, Brahm, and Benatti)\footnote{ESO programs 192.C-0224, 0102.C-0525, 0102.D-0483, 0101.C-0510, 0102.C-0451, 0103.C-0759} is labeled as "post". In total we obtained 103 observations from HARPS.
HARPS observations are processed on site using the standard ESO Data Reduction Software (DRS). The DRS pipeline data products were then reprocessed independently using the HARPS-TERRA code (Anglada-Escud\'e \& Butler 2012). HARPS-TERRA creates a high signal-to-noise template from the individual observations. Then, each observation is matched using a $\chi^2$ process relative to this template producing an RV for each observed spectrum. The code also computes stellar activity indices for the Calcium {\sc{ii}} H \& K lines ($\lambda_{H}$ = 3933.664 \AA, $\lambda_{K}$ = 3968.470 \AA), which are not available from the DRS-reduced spectra.  For the derivation of S-indices the code integrates the flux in these lines and compares with the flux on adjacent chunks in the continuum, following the procedure described in \citet{lovis11}. S-indices from HARPS-TERRA are calibrated to the Mt. Wilson system ($S_{MW}$) and they can be used as a direct proxy to monitor the chromospheric activity of the star. Additional activity indices such as the full-width at half-maximum of the cross-correlation function (CCF FWHM) and the bisector inverse slope (BIS) are taken directly from the fits headers of the DRS-reduced data products. Uncertainties in BIS are taken as twice the RV errors and the CCF FWHM uncertainties are 2.35 times the RV errors \citep{Zechmeister2018}.

\subsubsection{FEROS}
The FEROS spectrograph is mounted on the 2.2-m MPG/ESO telescope also at La Silla observatory and has a spectral resolution of R=48 000 \citep{FEROS}. Nine observations were made during September 10 and 19 under program 0103.A-9004(A) (PI: Vines) which were reduced with the \texttt{CERES} pipeline \citep{ceres}. After the reduction, \texttt{CERES} calculates RVs through the CCF method. We used a binary K5 mask for each epoch and fit a double Gaussian to the CCF in order to find the RV. A double Gaussian was used in order to account for scattered moonlight contamination. In addition to the RVs, \texttt{CERES} also provides the CCF FWHM and the BIS as activity indices. The uncertainties in BIS are calculated internally by \texttt{CERES} while the uncertainties of the CCF FWHM are calculated by dividing the standard deviation of the Gaussian fitted to the CCF, and dividing by the S/N at 5130 \AA, a procedure analogous to the RV uncertainty calculation done by \texttt{CERES}. S-indices index are computed from each 1D spectra after correcting to rest-frame following the procedure outlined by \cite{Jenkins2008} using the \texttt{Ceres-plusplus}\footnote{https://github.com/jvines/Ceres-plusplus} code.

\subsubsection{Minerva-Australis}
Minerva-Australis is an array of four PlaneWave CDK700 telescopes 
located in Queensland, Australia, fully dedicated to the precise 
RV follow-up of \tess{} candidates 
\citep[e.g.][]{toi677,toi257,AUMicBrett}.  The four telescopes can be 
simultaneously fiber-fed to a single KiwiSpec R4-100 high-resolution 
(R=80,000) spectrograph \citep{barnes12,addison19}.

HD 18599 was monitored by Minerva-Australis in its early commissioning 
period, with a single telescope between 2019 Jan 6 and 2019 Jan 29.  
Exposure times were 30 minutes, with some nights receiving two or three 
consecutive exposures.  We obtained a total of 31 observations, with a 
mean S/N of 23$\pm$5 per pixel.  RVs 
are derived for each telescope by cross-correlation, where the template 
being matched is the mean spectrum of each telescope.  The instrumental 
variations are corrected by using simultaneous Thorium-Argon arc lamp 
observations.

We show the RVs with the activity indicators in Table \ref{tab:rvs}.

\begin{table*}
	\centering
	\caption{RVs and Activity Indices for HD 18599}
	\label{tab:rvs}
	\begin{tabular}{cccccccccc}
BJD			&	RV  &  RV err  &  BIS		& BIS err &	S$_{\text MW}$ & S$_{\text MW}$ err & FWHM   & FWHM err & instrument	\\
(-2450000)	& (\ms)  &  (\ms)  &  (\ms)     & (\ms)   & (dex)          &      (dex)         & (\kms) & (\kms)   & \\
		\hline
6980.6959 &	12.7646  &	1.2472 &	5.2533 &  0.8099 & -0.0053 &  0.0028 & -0.0356 &  0.0029 &  HARPS\_pre \\
6980.7020 &	11.4832  &	1.2733 &	5.8890 &  0.8002 & -0.0034 &  0.0028 & -0.0337 &  0.0010 &  HARPS\_pre \\
6982.6788 &	16.0365  &	1.3771 &	8.2058 &  1.0231 &  0.0031 &  0.0033 &  0.0113 &  0.0032 &  HARPS\_pre \\
6982.6849 &	14.4035  &	1.9666 &	10.6088 &  1.0709 &  0.0214 &  0.0034 & 0.0067 &  0.0046 &  HARPS\_pre \\
6987.6309 &	-18.5854 &	1.9749 &	55.0689 &  1.1134 & -0.0055 &  0.0035 & 0.0504 &  0.0046 &  HARPS\_pre \\
6987.6371 &	-16.9837 &	1.8491 &	52.7322 &  1.1613 &  0.0077 &  0.0036 & 0.0564 &  0.0043 &  HARPS\_pre \\
6988.6928 &	-19.1186 &	1.2100 &	44.4527 &  0.8029 & -0.0180 &  0.0026 & -0.0556 &  0.0028 &  HARPS\_pre \\
\vdots    & \vdots   & \vdots  & \vdots     & \vdots  & \vdots  & \vdots  & \vdots  & \vdots  & \vdots \\
		\hline
	\end{tabular}
\end{table*}

%%%%%%%%%%%%%%%%%%%%%%%%%%%%%%%%%%%%%%%%%%%%%%%%%%%%%%%%%%%%%%%%%%%%%%%%%%%%%%%%%%%%%%%%%%%%%%%%%%%%%%%%%%%%%%%%%%%%%%%%%%%%%%%%%%%%%%%%%%%%%%%%%%%%%%%%%%%%%%%%%%

\section{Analysis}
\label{sec:analysis}

%%%%%%%%%%%%%%%%%%%%%%%%%%%%%%%%%%%%%%%%%%%%%%%%%%%%%%%%%%%%%%%%%%%%%%%%%%%%%%%%%

\subsection{Stellar Properties}
\label{sub:stellar}
\subsubsection{\species{}}
We analysed the 1D stacked HARPS spectra using \species{} \citep{species, SPECIESII}. \species{} is an automated code to derive stellar parameters for large samples of stars, using high resolution echelle spectra.
It makes use of equivalent widths from a number of neutral and ionized iron lines \citep[measured with EWComputation,][]{SPECIESII} to derive the atmospheric parameters (temperature, metallicity, surface gravity, and microturbulence). Together with ATLAS9 model atmospheres \citep{Castelli2004}, it solves the radiative transfer and hydrostatic equilibrium equations using MOOG \citep{sneden_moog}, imposing local thermodynamic equilibrium (LTE) conditions, as well as excitation and ionization equilibrium. The correct atmospheric parameters are found when the abundance for ionized and neutral iron are the same, the obtained iron abundance is the same as the one used to create the model atmosphere, and there is no correlation between the neutral iron abundances with the excitation potential, and with the reduced equivalent widths ($W/\lambda$). Finally, rotational and macro turbulent velocity were derived using spectral line fitting and analytic relations, respectively. \species{} finds an effective temperature of 5109$\pm$50~K, a $\log~g$ of 4.41$\pm$0.07~dex, a [Fe/H] of -0.01$\pm$0.05~dex and a vsini of 5.2$\pm$0.2~km~s$^{-1}$.

\subsubsection{\ariadne{}}

We used the temperature, $\log~g$ and [Fe/H] found by \species{} as Gaussian priors on a Spectral Energy Distribution (SED) analysis performed with \ariadne{}\footnote{https://github.com/jvines/astroARIADNE} \citep{ariadne} to find the final set of bulk stellar parameters. \ariadne{} is a publicly available python tool designed to fit catalog photometry to different stellar atmosphere models grids, \texttt{Phoenix V2} \citep{Husser2013}; \texttt{BT-Settl}, \texttt{BT-Cond}, \texttt{BT-NextGen} \citep{Hauschildt1999, Allard2012}, \cite{Castelli2004} and \cite{Kurucz}; which have been convolved with the following filter response functions:

\begin{itemize}
    \item Johnson UBV
    \item Tycho-2 BtVt
    \item 2MASS JHK$_\text{s}$
    \item SDSS \textit{ugriz}
    \item Gaia DR2v2 G, RP and BP
    \item GALEX NUV and FUV
    \item TESS
\end{itemize}

The SEDs are modelled by interpolating the model grids in T$_{\rm eff}$-$\log~g$-[Fe/H] space, scaling the synthetic flux by $(R/D)^2$ and accounting for interstellar extinction through the extinction in the V band, A$_{\rm V}$. Additionally we include an excess noise parameter for every photometric observation in order to account for underestimated uncertainties. \ariadne{} samples the posterior space using the \texttt{dynesty} \citep{dynesty} implementation of (dynamic) nested sampling \citep{dynamicnestedsampling}. Nested Sampling \citep{nestedsampling1, nestedsampling2} algorithms are designed to estimate the Bayesian evidence of a model producing posterior distributions as a by-product.

We modeled the SED with four models, Phoenix V2, BT-Settl, \cite{Castelli2004} and \cite{Kurucz}, and finally averaged the posterior parameters from each model, weighting them by their respective relative probability that was computed from their Bayesian evidence estimates, resulting in precise parameters that take into account the different micro-physics and geometry of the different models.

Additional priors for distance, radius, and A$_{\rm v}$ were drawn from the Bailer-Jones distance estimate from Gaia DR2 \citep{Bailer-Jones2018}, a uniform prior from 0.5 to 20 R$_\odot$ and a uniform prior from 0 to 0.038, the maximum line-of-sight extinction from the re-calibrated SFD galactic dust map \citep{Schlegel1998, Schlafly2011}.

\begin{figure}
	\includegraphics[width=\columnwidth]{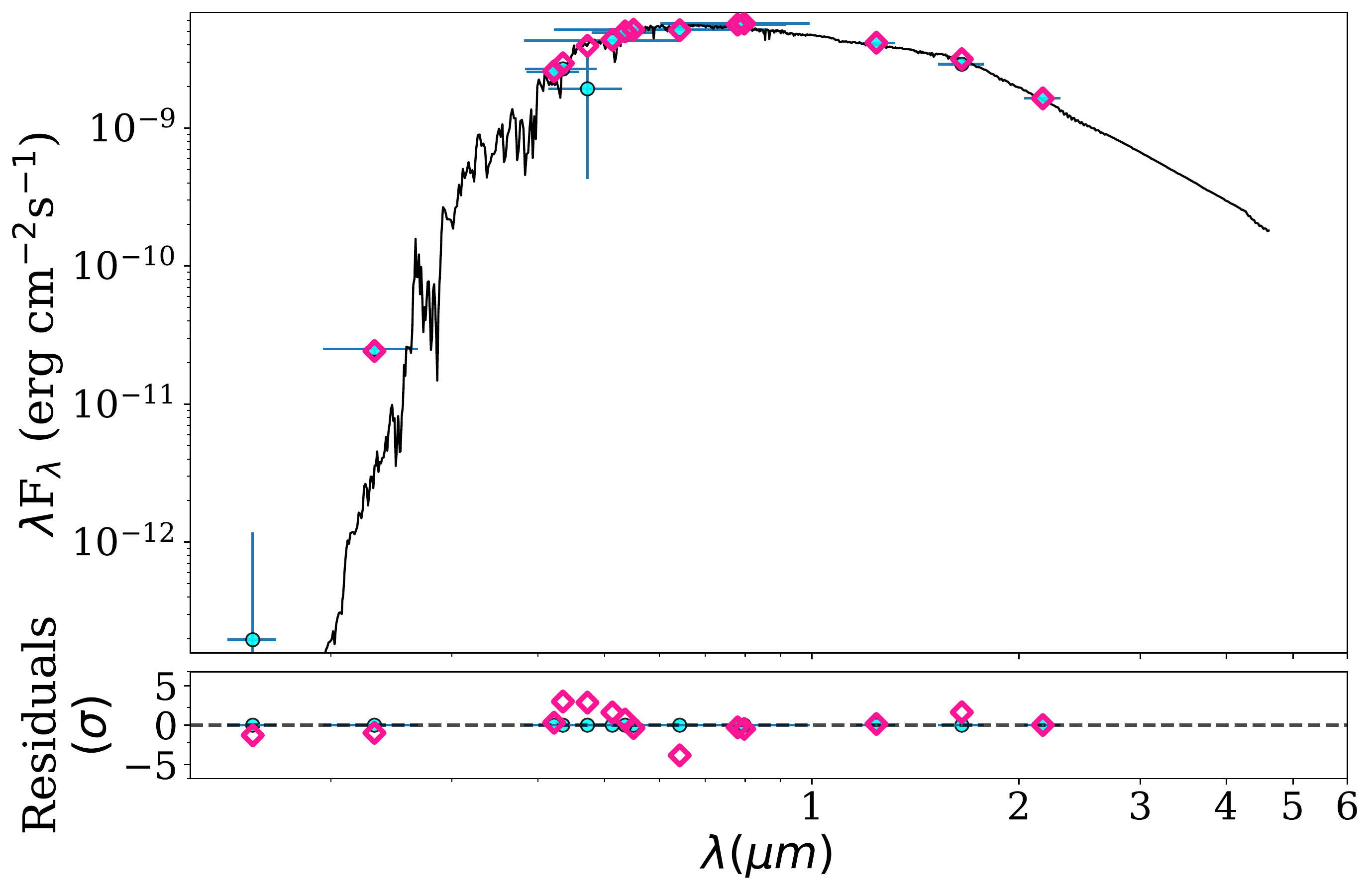}
    \caption{The best fitting Castelli \& Kurucz SED (black line) for HD 18599 based on the photometric data (cyan points) presented in Table~\ref{tab:stellar} is shown in the top panel.  Horizontal errorbars show the bandpass width. The pink diamonds show the synthetic magnitudes at the wavelengths of the photometric data.  The lower panel shows the residuals to the best fit model.}
    \label{fig:SED}
\end{figure}

In Figure \ref{fig:SED} we show the SED for \starname\, and in Table \ref{tab:stellar} we report the relevant observational properties with our derived properties, along with the method used.

HD 18599 is previously known to be a young field star based on previous studies \citep{Grandjean20, Grandjean21}. In a more recent study, de Leon et al. (submitted) employed various methods to estimate the age of HD 18599. In that paper, we measured an equivalent width (EW) of 42$\pm$2m\AA\,for Li I $\lambda$6708. This measurement generally agrees with those found in stars in Hyades and Praesepe corresponding to an age of $\sim$800 Myr. We also estimated the stellar age by taking advantage of the observed chromospheric activity together with empirical age-activity-rotation relations. In particular, we used the chromospheric activity indicator, $\log R'_{HK} = -4.41 \pm 0.02$ from \citealt{boro18} which predicts an age of $0.30 \pm 0.05$~Gyr. Whilst this star is an X-ray source based on detection from ROSAT, the X-ray strength is weak ($\log Lx/Lbol=-4.64\pm0.25$) which corresponds to $1-\sigma$ age range from X-ray of $475^{+734}_{-305}$ Myr. 

HD 18599 exhibits significant spot-modulated rotational signals in the four sectors of TESS observations. Using a periodogram analysis, we measured the rotation period of HD 18599 for each of these sectors of observations, finding that it has a mean rotation period of $8.71\pm0.31$~d. We compute a gyrochronological age estimates of 386 Myr, with $3-\sigma$ range of 261-589 Myr based on \citealt{Mamajek2008} model, whereas we compute an age of 247 Myr, with 3-$\sigma$ range of 185-329 Myr based on \citealt{barnes07} model. We can further corroborate the activity-based age estimate by also using empirical relations to predict the stellar rotation period from the activity. For example, the empirical relation between $R'_{HK}$ and rotation period from \citealt{Mamajek2008} predicts a rotation period for this star of $9.7 \pm 1.3$~d, which is compatible with the rotation periods above as well as with the rotation period of 8.69~d reported by KELT.

\begin{figure}
	\includegraphics[width=\columnwidth]{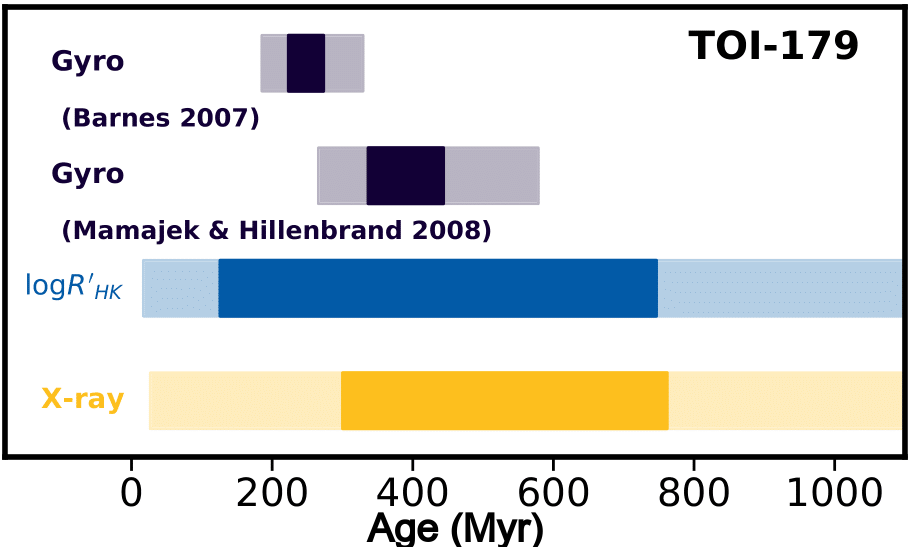}
    \caption{Different age estimates for HD 18599 using various methods.}
    \label{fig:age}
\end{figure}

Figure \ref{fig:age} summarizes the various age estimates using the different methods discussed in de Leon et al (submitted) which is consistent with the age between 0.1 to 1~Gyr. For simplicity, we adopt a median age of 300 Myr throughout the rest of the paper.

An asteroseismology study of \starname\, was also performed, finding that the active nature of the star produces photometric variability much larger than the amplitude of the expected oscillation modes, making the visual detection of the oscillations very tough. The details of the analysis and the results can be found in  Samadi-Ghadim et al. (2022 in prep.).

\begin{table}
	\centering
	\caption{Stellar Properties for HD 18599}
	\begin{tabular}{lcc} % six columns, alignment for each
	Property	&	Value		&Source\\
	\hline
    \multicolumn{3}{l}{Astrometric Properties}\\
    R.A.		&	\mbox{$02^{\rmn{h}} 57^{\rmn{m}} 02\fs8835$}			&Gaia	\\
	Dec			&	\mbox{$-56\degr 11\arcmin 30\farcs 7297$}			&Gaia	\\
    2MASS I.D.	& J02570294-5611314	&2MASS	\\
    TIC         & 207141131         &\tess{} \\
    Gaia DR2 I.D. & 4728513943538448512 &Gaia\\
    $\mu_{{\rm R.A.}}$ (\masy) & -36.676$\pm$0.042 & Gaia \\
	$\mu_{{\rm Dec.}}$ (\masy) & 50.599$\pm$0.045 & Gaia \\
	$\varpi$ (mas)  & 25.9011$\pm$0.0244    &Gaia\\
    \\
    \multicolumn{3}{l}{Photometric Properties}\\
	V (mag)		&8.968$\pm$0.012 	&APASS\\
	B (mag)		&10.015$\pm$0.004		&APASS\\
	g (mag)		&10.20$\pm$0.05		&APASS\\
	V$_\text{T}$ (mag)  &9.084$\pm$0.016    &Tycho2\\
	B$_\text{T}$ (mag)  &10.015$\pm$0.004 &Tycho2\\
    G (mag)		&8.7312$\pm$0.0007	&Gaia\\
    RP (mag)    &8.1384$\pm$0.0019    &Gaia\\
    BP (mag)    &9.2113$\pm$0.0020    &Gaia\\
    \tess{} (mag)	&8.1796$\pm$0.0060		&\tess{}\\
    J (mag)		&7.428$\pm$0.018		&2MASS	\\
   	H (mag)		&7.029$\pm$0.015		&2MASS	\\
	Ks (mag)	&6.883$\pm$0.020		&2MASS	\\
    NUV (mag)	&15.684$\pm$0.012		&GALEX	\\
    FUV (mag)	&21.389$\pm$0.264	&GALEX	\\
    \\
    \multicolumn{3}{l}{Derived Properties}\\
    T$_{\rm eff}$ (K)    & 5083$\pm$23               &\ariadne{}\\
    $\left[{\rm Fe}/{\rm H}\right]$	 & -0.05$\pm$0.04		&\ariadne{}\\
    vsini (\kms)	     &	5.2$\pm$0.2			    &\texttt{SPECIES}\\
    log g                &	4.40$\pm$0.07	&\ariadne{}\\\vspace{3pt}
    \mstar (\msun) & 0.807$^{+0.019}_{-0.007}$		        &\ariadne{}\\\vspace{3pt}
    \rstar (\rsun) & 0.798$^{+0.006}_{-0.007}$	            &\ariadne{}\\\vspace{3pt}
    $\rho$ (\gccc) & 2.241$^{+0.081}_{-0.077}$              & This work \\\vspace{3pt}
    Age	(Myr)			& 300				&Gyrochronology\\\vspace{3pt}
    Distance (pc)	&  38.585$^{+0.110}_{-0.150}$	                &\ariadne{}\\\vspace{3pt}
    Av (mag)    & 0.030$^{+0.002}_{-0.014}$    & \ariadne{} \\\vspace{3pt}
    $P_{\rm rot}$  &  8.74$\pm$0.05  & This work \\
	\hline
    \multicolumn{3}{l}{2MASS \citep{2MASS}; UCAC4 \citep{UCAC};}\\
    \multicolumn{3}{l}{APASS \citep{APASS}; WISE \citep{WISE};}\\
    \multicolumn{3}{l}{Gaia \citep{GAIA, GAIA_DR2}; GALEX \citep{GALEXDR5};}\\
    \multicolumn{3}{l}{TESS \citep{TESS_CAT}} \\
    \multicolumn{3}{l}{$q_{1,2}$ are the Kipping Limb Darkening parameters.}
	\end{tabular}
    \label{tab:stellar}
\end{table}

%%%%%%%%%%%%%%%%%%%%%%%%%%%%%%%%%%%%%%%%%%%%%%%%%%%%%%%%%%%%%%%%%%%%%%%%%%%%%%%%%

\subsection{Stellar Rotation}
\label{sub:rotation}

From {\it Kepler} data it has been shown that the analysis of photometric time series can yield good estimates of the stellar rotation period (see, e.g. \citealt{McQuillan2013, Giles2017}). Rotationally modulated spots on the stellar surface can produce a periodic signal that can be detected in the photometry, since these spots are regions of diminished flux. Additionally, rotational modulations can be a source of false-positive exoplanets signatures, for which the \tess{} photometry, combined with ground-based data sources, can serve as a powerful diagnostic for assessing the properties of known exoplanets \citep{Kane2021,Simpson2022}.

\begin{figure}
	\includegraphics[width=\columnwidth,angle=0]{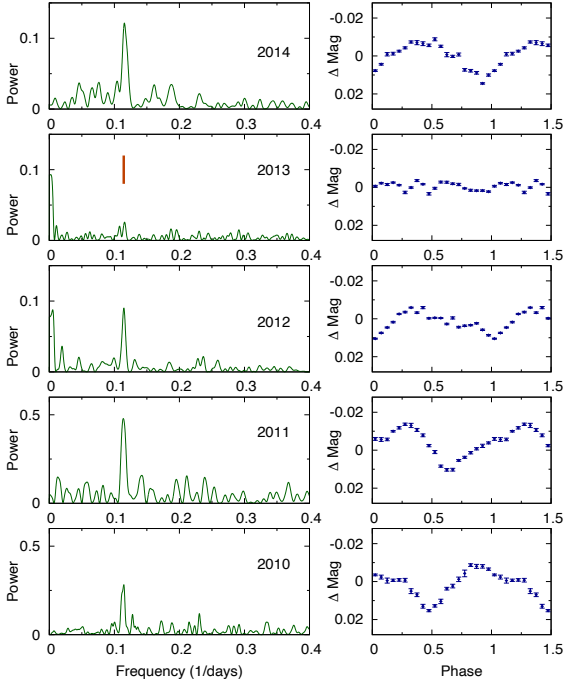}
    \caption{Periodograms of the WASP-South data for HD 18599 (left) along with the modulation profiles from folding the data (right). The red line is at a period of 8.74 days.}
    \label{fig:wasp}
\end{figure}

We searched each season's WASP light curve for rotational modulations using the methods presented in \citet{2011PASP..123..547M}. There is a clear and persistent modulation seen in 4 of the five seasons (Figure~\ref{fig:wasp}). The mean period is 8.74 $\pm$ 0.06 d, the amplitude ranges from 6 to 10 mmag, and the false-alarm likelihoods are less than 10$^{-3}$.  In 2013 the modulation is marginally present, but with a much lower amplitude of 1--2 mmag. 

We also computed the Autocorrelation Function \citep[ACF,][]{McQuillan2013} for each \tess{} light curve, using the PCDSAP corrected data provided by the SPOC in order to find a possible value for the stellar rotation period. We follow the procedure described in \citet{Soto2018}, where we model the ACF as an underdamped Simple Harmonic Oscillator (uSHO) \citep{LopezMorales2016}. We define

\begin{equation}
y(t) = e^{-t/\tau_{\rm AR}} \left[ A \cos \left( \frac{2\pi t}{P_{\rm rot}}\right) + B \cos \left(\frac{4\pi t}{P_{\rm rot}} \right) \right] + y_{0}
\end{equation}

Where $\tau_{\rm AR}$ is the correlation time-scale and $P_{\rm rot}$ is the rotation period. 

We estimated $P_{\rm rot}$ using the Edelson \& Krolik method \citep{EdelsonKrolik1988} implemented in \texttt{astroML} Python package via a least squares fitting process \citep{astroML}. Results are tabulated in Table \ref{tab:ACF}.

\begin{table}
	\centering
	\caption{ACF results from the \tess{} light curve.}
	\label{tab:ACF}
	\begin{tabular}{ccc}
	Sector	&	$P_{\rm rot}$        	& $P_{\rm rot}$ err\\
    	&	(days)	& (days)\\
	\hline
    2 & 4.522 & 0.001 \\
    3 & 4.526 & 0.002 \\
    29 & 8.605 & 0.281 \\
    30 & 8.791 & 0.110 \\
	\hline
	\end{tabular}
\end{table}

The rotation period from the first two sectors are consistent with each other and consistent with being an alias from the rotation period derived from WASP data. This is further confirmed with the estimated rotation period of the latter sectors being consistent with the one derived by WASP as well.

\begin{figure}
	\includegraphics[width=\columnwidth,angle=0]{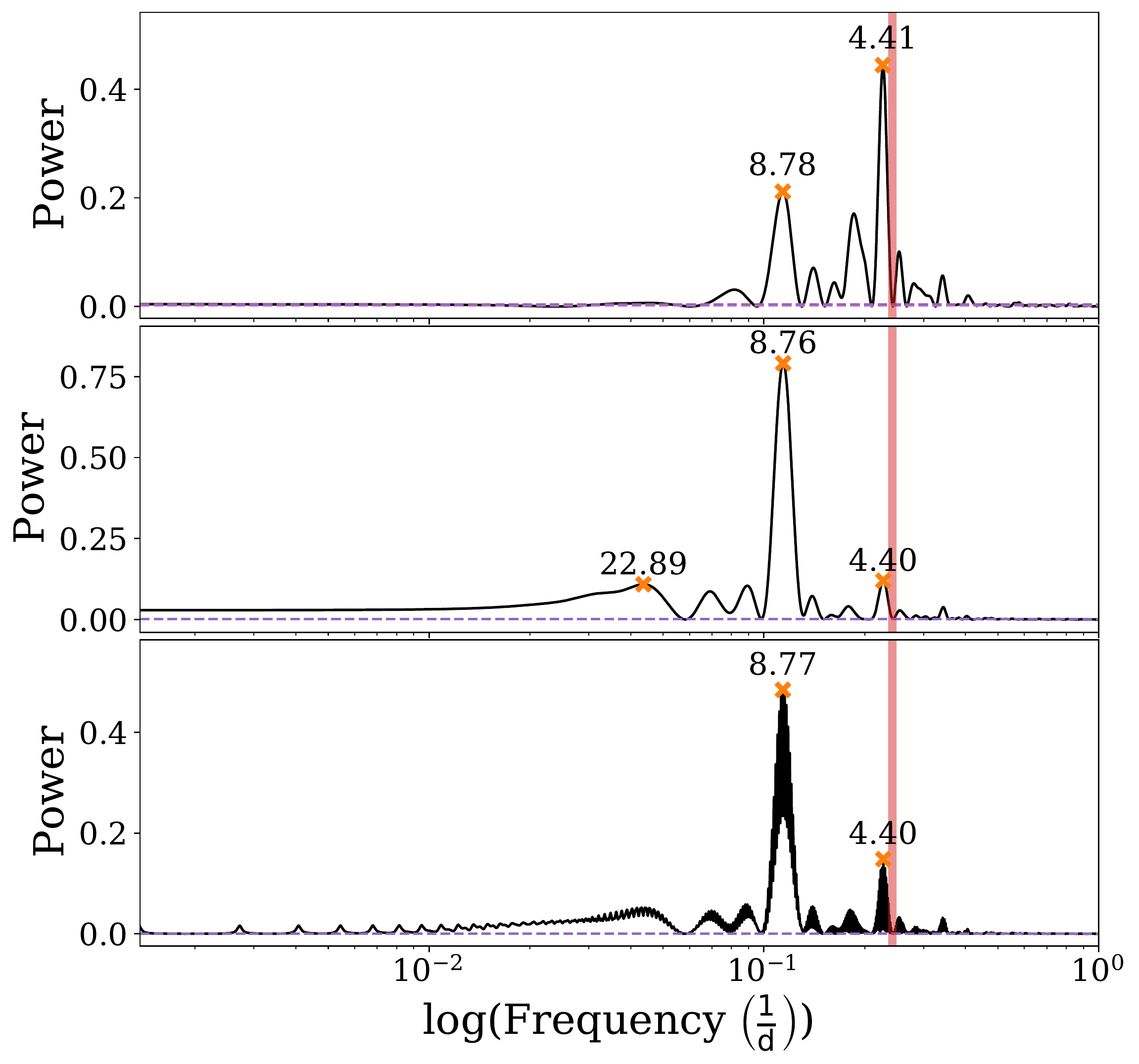}
    \caption{From top to bottom: the GLS periodogram for the \tess{} light curves from sectors 2 and 3 (top), sector 29 and 30 (middle), and the three sectors combined (bottom). The orange crosses show the most significant peaks with their respective periods labeled on top. The vertical red line marks the 4.1375-day planet candidate from the \tess{} data. The dashed horizontal line are the false alarm probabilities for 0.1, 5 and 10\%, they turn out to be superimposed.}
    \label{fig:tess_gls}
\end{figure}

We ran the Generalized Lomb-Scargle periodogram \citep[GLS,][]{Zechmeister2009} in a grid of 50,000 trial frequencies ranging from one day to two years in period space on the \tess{} data to find the possible sinusoidal signals. We show these in Figure \ref{fig:tess_gls} where the dashed line represent the 0.1, 5, and 10\% false-alarm probabilities (FAP), but they happen to coincide due to the amplitude of the oscillations. The top panel of Figure \ref{fig:tess_gls} shows the GLS periodogram of sectors 2 and 3, where the two most strong signals correspond to 4.41 and 8.78 days. The middle panel shows the periodogram of sectors 29 and 30 where the most dominant signal is 8.76 days, the second most strong signal is 22.89 days. We show a third signal, which while not as significant, corresponds to the same period as the most strong signal found in sectors 2 and 3, 4.40 days. The bottom panel shows the periodogram of the combined three sectors where the most dominant signal is the 8.77 day one and the 4.40 day signal is the second most significant one.

\begin{figure*}
	\includegraphics[width=2\columnwidth,angle=0]{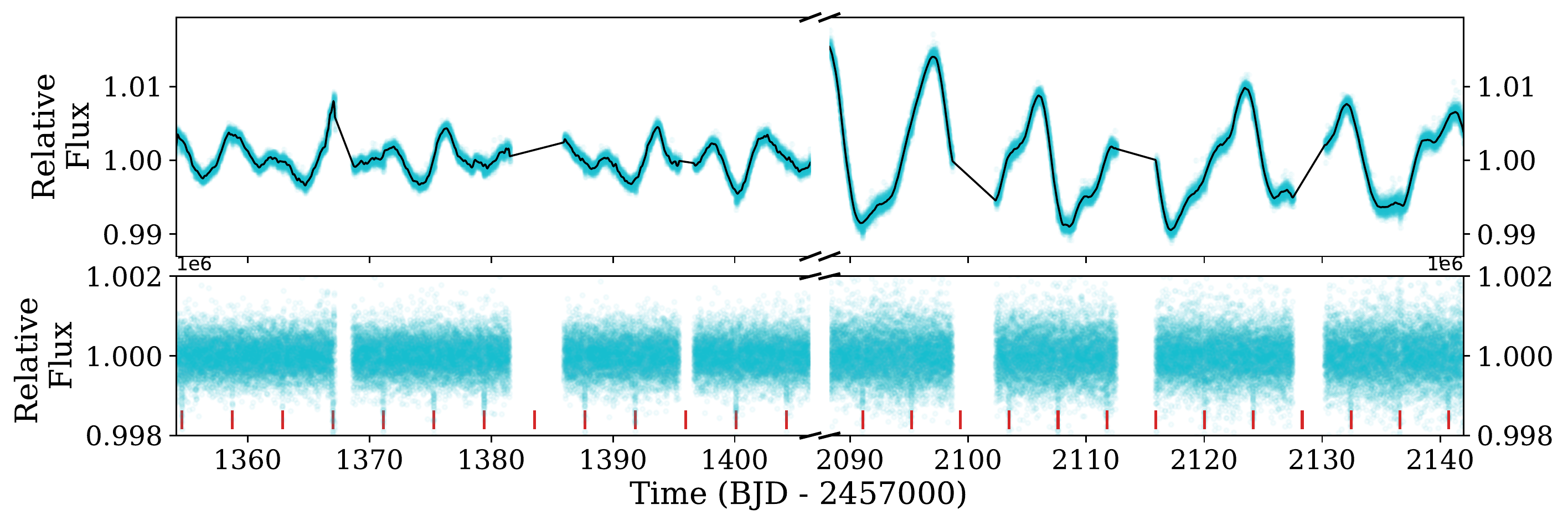}
    \caption{The normalized \tess{} light curve including data from the three sectors, with the best fitting SHO kernel GP (black line). The bottom panel shows the detrended light curve in ppm. The vertical red lines show the transit features.}
    \label{fig:detrend}
\end{figure*}

Using \texttt{Juliet}\footnote{https://juliet.readthedocs.io/en/latest/} \citep{juliet} and \texttt{MultiNest} via the Python wrapper \texttt{pyMultinest} \citep{multinest, pymultinest}, we fit a Gaussian Process (GP) using a Simple Harmonic Oscillator (SHO) kernel from \textit{celerite} \citep{celerite} to the out-of-transit \tess{} light curve in order to remove the stellar activity signal present. We used the rotation period derived from WASP as a Gaussian prior for the characteristic frequency of the SHO. We found the best fitting parameters to be $S_0 = 0.0008\pm0.0001$, $Q = 0.11\pm0.01$ and $\omega_0 = 0.719\pm0.004$. A quality factor less than $1/2$ means the system is overdamped and thus there is no oscillation aside from the rotation of the star, which is captured by the GP as seen in the top panel of Figure \ref{fig:detrend}. These parameters translate to a rotation period of $8.74\pm0.05$ days. The top panel of Figure \ref{fig:detrend} shows the raw light curve with the best fitting GP, and the bottom panel shows the detrended light curve, with the red vertical lines highlighting the transit features. 

\subsection{Activity Indices}
\label{sec:activity}

We searched for correlations between the RVs and the activity indices: (BIS), CCF FWHM and S$_{\rm MW}$, and we found a strong anticorrelation between the BIS and RVs, no correlation for the CCF FWHM, and a weak correlation for  S$_{\rm MW}$, quantified by the Pearson $r$ correlation coefficients: -0.62, 0.05, and 0.49, respectively. We show our thresholds for correlation in Table \ref{tab:pearsonr}.

\begin{table}
	\centering
	\caption{Adopted thresholds for the Pearson $r$ correlation coefficient. We only show the positive values, but the same thresholds apply for their negative counterparts.}
	\label{tab:pearsonr}
	\begin{tabular}{cc}
	$r$	&	Strength \\
	\hline
0.00 to 0.19 & Very weak\\
0.20 to 0.39 & Weak\\
0.40 to 0.59 & Moderate\\
0.60 to 0.79 & Strong\\
0.80 to 1.00 & Very strong\\
	\hline
	\end{tabular}
\end{table}

\begin{figure*}
	\includegraphics[width=2\columnwidth,angle=0]{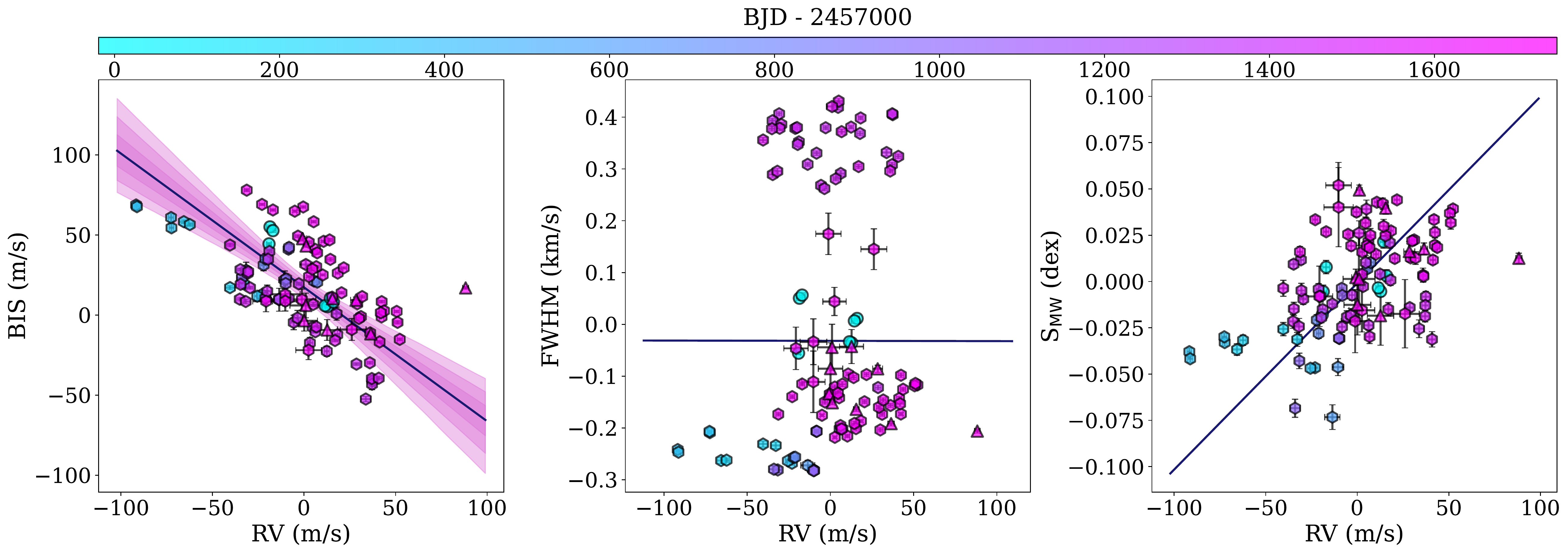}
    \caption{From left to right: The bisector velocity span with 1, 2 and 3 sigma confidence regions, FWHM, and S-index correlation with the RVs after subtracting their respective mean values. HARPS\_pre, HARPS\_post, and FEROS are denoted by circles, hexagons and triangles respectively. Each observation is color coded by its BJD timestamp. In all cases the blue line corresponds to the best fitting linear model. Confidence regions were not added for the FWHM and S index panels due to them being unconstrained.}
    \label{fig:corrs}
\end{figure*}

In addition to the Pearson $r$ coefficients, we fit a linear model to each of the correlations using Markov Chain Monte Carlo (MCMC) through \texttt{emcee} \citep{emcee} to probe the posterior parameter space. We employed a gaussian likelihood that takes into account uncertainties in both axes. We show the best fitting model parameters in Table \ref{tab:corrs} and the correlations in Figure \ref{fig:corrs}. Given that the slopes of the CCF FWHM and S$_{\rm MW}$ are statistically consistent with zero, we conclude that the only significant correlation is the BIS.

\begin{table}
	\centering
	\caption{Intercept and slopes of the best fitting linear models of each correlation.}
	\label{tab:corrs}
	\begin{tabular}{ccc}
	Correlation	&	Slope & Intercept \\
	\hline\vspace{3pt}
BIS & -0.836$^{+0.088}_{-0.096}$ & 17.302$^{+2.222}_{-2.254}$ \\\vspace{3pt}
CCF FWHM & 0.000$^{+0.005}_{-0.005}$ & -0.032$^{+0.161}_{-0.162}$ \\\vspace{3pt}
S$_{\rm MW}$ & 0.001$^{+0.005}_{-0.005}$ & -0.001$^{+0.162}_{-0.162}$ \\
	\hline
	\end{tabular}
\end{table}

Using the GLS periodogram we searched for periodicities in the RVs, BIS, S$_{\rm MW}$ indices, and CCF FWHM. Signals at 1, 0.5 and 0.3 days arise from the window function, as well as a broad signal around 800 to 1000 days which reappears in the RVs, S$_{\rm MW}$, and the FWHM. The BIS shows a peak at 4.40 days which coincides with the Prot alias from \tess{} and WASP. Given that the period candidate is very close to the rotation period alias tracked by the BIS, a straightforward decorrelation is not possible since it would also inadvertently remove the candidate signal from the data, therefore, we chose to model the raw RVs but with the incorporation of different noise and activity models.

\begin{figure}
	\includegraphics[width=\columnwidth,angle=0]{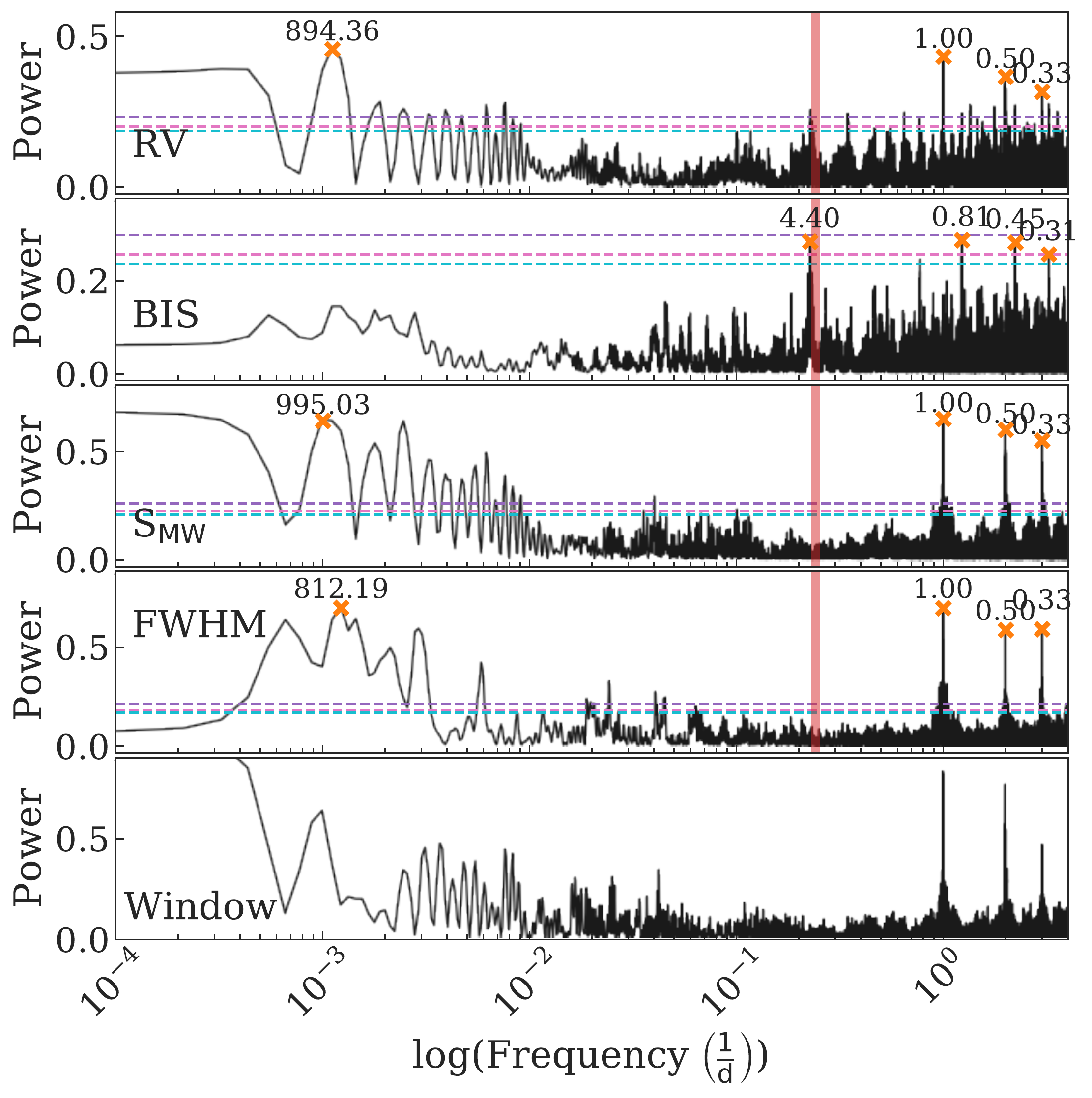}
    \caption{From top to bottom: the GLS periodogram for the RVs and activity indices from HARPS and FEROS: BIS, S-index, and CCF FWHM. The bottom panel shows the window function. Horizontal lines from top to bottom are the 0.1, 5 and 10\% false alarm probabilities. The vertical red line marks the 4.1375-day planet candidate signal from the \tess{} data. The orange crosses show the most relevant peaks with their values in period written above them.}
    \label{fig:all_gls}
\end{figure}

%%%%%%%%%%%%%%%%%%%%%%%%%%%%%%%%%%%%%%%%%%%%%%%%%%%%%%%%%%%%%%%%%%%%%%%%%%%%%%%%%

\subsection{Global Modeling}
\label{sub:global}

We jointly modeled the detrended \tess{} data with the HARPS and FEROS RVs, in order to obtain the mass, radius, and orbital parameters of TOI-179 b, with \texttt{EMPEROR.T}, an upgraded version of \texttt{EMPEROR} (Pena \& Jenkins in prep) that handles both RVs and light curves.

\texttt{EMPEROR} models transits with the transit light curve calculation code \texttt{PyTransit} \citep{PyTransit} using the quadratic limb darkening law from \citet{quadld}, along with the triangular limb-darkening parametrization proposed by \citet{Kipping2016}. The transit model also includes an offset and white noise term, as well as a dilution factor \citep{juliet}, which we fixed to unity due to the expected dilution being $<1\%$ (de Leon et al. 2022, in prep.).
RVs are modeled with the Keplerian model including a velocity offset $\gamma$ and an instrumental white noise parameter $\sigma$ for each instrument and a first order acceleration term. \texttt{EMPEROR} implements instrumental $q$ order autoregressive and $p$ order moving average models \citep[AR($q$)MA($p$);][]{Mikko} as a correlated noise model that deals with noise from, for example, stellar variability. Additionally, \texttt{EMPEROR} has incorporated a linear stellar activity correlation model following the description in \citet{Anglada-Escude2016}.  \texttt{EMPEROR} also allows for the modeling of the stellar density, which has been shown to improve parameter estimation as it deals with the degeneracy between the radius ratio and the impact parameter \citep{Sandford2019, Vines2019}. The algorithm assumes a Gaussian generative model and thus both the RV jitter and photometric white noise are added in quadrature to their respective observational uncertainties. The described model is given by

\begin{equation}
    rv_{i, \rm{ins}} = k(t_i) + ar_{i, \rm{ins}}(q) + ma_{i, \rm{ins}}(p) + A_{i, \rm{ins}} + \gamma_{\rm ins} + \dot{\gamma}_{\rm ins} + \sigma_{i, \rm ins}
\end{equation}

\begin{equation}
    ar_{i, \rm{ins}}(q) = \sum_{j=1}^q\phi_{j, \rm ins}\exp\left(\frac{t_{i-j} - t_i}{\alpha_{j, \rm ins}}\right)rv_{i-j, \rm ins}
\end{equation}

\begin{equation}
    ma_{i, \rm{ins}}(p) = \sum_{j=1}^q\omega_{j, \rm ins}\exp\left(\frac{t_{i-j} - t_i}{\beta_{j, \rm ins}}\right)\epsilon_{i-j, \rm ins}
\end{equation}

\begin{equation}
    A_{i, \rm ins} = \sum_\xi C_{\xi, \rm ins}\xi_{i, \rm ins}
\end{equation}

Where $k(t_i)$ is the Keplerian model evaluated at time $t_i$, $\dot{\gamma}_{\rm ins}$ is an acceleration term for each instrument, $\gamma_{\rm ins}$ is the velocity offset of instrument ins, $ar_{i, \rm{ins}}(q)$ and $ma_{i, \rm{ins}}(p)$ are the $q$th and $p$th order autoregression and moving average respectively, with $\phi$ and $\omega$ being the strength of the correlated noise, $\alpha$ and $\beta$ are the characteristic time associated to the correlated noise, and $\epsilon_{i-j, \rm ins}$ are the residuals of the model so far. $A_{i, \rm{ins}}$ is the linear activity indices correlation model where $\xi$ denotes the activity indices for each instrument and $C_{\xi, \rm ins}$ are the activity coefficients. Finally $\sigma_{i, \rm ins}$ is the white noise term added for each instrument.

Sampling is carried out through \texttt{emcee} version 2.2.1 and its Parallel Tempering MCMC (PTMCMC) module, which is capable of sampling multimodal phase spaces without getting stuck in local maxima. This is achieved by modifying the posterior space as shown in Equation \ref{eq:tempered}, where $\beta = 1/T$ with $T$ a temperature, $\mathcal{L}$ the original likelihood, and $\Pi$ is the prior density. This effectively flattens the posterior space with increasing temperature, allowing high temperature chains to sample the posterior more easily.

\begin{equation}
    \label{eq:tempered}
    \mathcal{L}_\beta(x) = \mathcal{L}^\beta(x)\Pi(x)
\end{equation}

Given the active nature of TOI-179 and the significant amount of scatter in the data, we decided to test six different configurations for modeling the system: an AR(1)MA(1) model; activity correlations with the BIS; AR(1)MA(1) plus activities; AR(1)MA(0) model; AR(0)MA(1) model; and a pure keplerian model, hereafter we will refer to those as runs 1, 2, 3, 4, 5, and 6. Additionally we ran a noise model with activity correlations as a baseline for the model comparison. For the sampling we used 5 temperatures, 1500 walkers and 6000 steps, with a burn in period of half the number of steps, totaling 45 million samples for the search phase, and 1500 walkers with 6000 steps for the sampling stage, ending the fitting process with 9 million samples for parameter inference. We chose the temperature ladder to decrease as $1/\sqrt{5}^i$ where $i=0, 1, 2,... T$ and $T$ the number of temperatures minus one. We used the same configuration for the walkers, temperatures and priors for all configurations mentioned previously.

\subsubsection{Prior Selection}
\label{sec:priors}

We chose the period prior to be a Jeffreys prior due to it being uninformative in period space, and we chose the bounds to be from 0.1 to 6 days in order to avoid possible aliases at 8 days. We applied the transformation $s = \sqrt{e}\sin(\omega)$, $c = \sqrt{e}\cos(\omega)$, both bounded from -1 and 1 to fit for the eccentricity, $e$, and argument of periastron, $\omega$. At each step of the MCMC sampling \texttt{EMPEROR} makes sure $e$ and $\omega$ are physically plausible, i.e. $0\leq e<1$ and $0\leq\omega\leq2\pi$. Additionally, an external prior is applied to the eccentricity in order to penalize high values while still allowing them if the data argues for it. The rest of the planetary parameters have uninformative, flat priors. For the stellar density we chose a prior drawn from the mass and radius estimate derived with \ariadne{} using the symmetrized 3$\sigma$ credible interval values as the standard deviation to allow for a more conservative prior. We chose uninformative priors for the instrumental parameters as well, with the exception of the RV jitter, where a Normal prior was chosen in order to penalize high values, and the photometric offset, which also has a Normal prior centered around zero since there is no significant offset in the detrended light curve. We summarise the priors in Table \ref{tab:priors}.

\begin{table}\label{tab:priors}
\caption{Prior choices used in this work}
\centering
\begin{tabular}{ll}
\hline
Parameter                         & Prior                   \\ \hline
Orbital Parameters & \\ \hline
P [days]  & $\ln \mathcal{U}(0.1, 6)$  \\
K [ms$^{-1}$]         & $\mathcal{U}(0, 50)$ \\ 
T$_c$ [JD]           & $\mathcal{U}(min(t^\dagger), max(t))$       \\ 
$\sqrt{e}\sin(\omega)$ & $\mathcal{U}(-1, 1)$ \\ 
$\sqrt{e}\cos(\omega)$ & $\mathcal{U}(-1, 1)$ \\ 
$\omega$ [rads]         & $\mathcal{U}(0, 2\pi)$      \\ 
e                 & $\mathcal{N}(0, 0.3^2)$            \\ 
$R_p/R_{*}$  &  $\mathcal{U}(0.01, 0.5)$ \\
b  & $\mathcal{U}(0, 1)$  \\ \hline
Stellar Parameters & \\ \hline
$\rho_{*}$ [gcm$^{-3}$]  &  $\mathcal{N}(2.241, 0.479^2)$ \\
$C^\ddagger_\xi$  & $\mathcal{U}(-max(C_\xi), max(C_\upsilon))$ \\
$q^{\textasteriskcentered}_1$  &  $\mathcal{U}(0, 1)$ \\
$q^{\textasteriskcentered}_2$  &  $\mathcal{U}(0, 1)$ \\ \hline
% &&
RV Noise Parameters                  &                                    \\ \hline
$\gamma$ [ms$^{-1}$]          & $\mathcal{U}(0, 3max(|rv|))$  \\  
$\sigma$ [ms$^{-1}$]    & $\mathcal{N}(5, 5^2)$       \\
MA Coefficient $\omega$        & $\mathcal{U}(-1, 1)$       \\ 
MA Timescale $\beta$ [days]        & $\mathcal{U}(0, 10)$       \\
AR Coefficient $\phi$        & $\mathcal{U}(-1, 1)$       \\ 
AR Timescale $\alpha$ [days]        & $\mathcal{U}(0, 10)$       \\ \hline
Transit Noise Parameters & \\ \hline
offset [ppm]  & $\mathcal{N}(0, 0.1^2)$ \\
jitter [ppm]  & $\ln{\mathcal{U}}(0.1, 10000)$ \\
dilution  & fixed $(1)$ \\ \hline
% &&
Acceleration Parameter            &       \\     \hline                     
$\dot{\gamma}$ [ms$^{-1}$/yr]   & $\mathcal{U}(-1, 1)$    \\
\hline
\end{tabular}\\
$^\dagger t=$Time baseline of \tess{} data.\\
$^\ddagger$ Activity indices were mean subtracted and normalized to their RMS. \\
$^{\textasteriskcentered}$ Kipping LD parameters.
\end{table}

\subsection{Model Selection}

We use the log posterior probability, Bayes Information Criterion (BIC) and Akaike Information Criterion (AIC), defined in Equations \ref{eq:BIC} and \ref{eq:AIC} respectively, where $n$ is the number of data points, $k$ the number of parameters, and $\mathcal{L}$ the maximum likelihood of the model, to do model comparison, using the log posterior to select the final model. The most probable model is Run 2, followed by Run 6, which suggests that the data is dominated by white noise instead of correlated noise. The semiamplitude of all runs is consistent, with the exception of Run 1, which produces a smaller semiamplitude by a factor of two, but still within the 3$\sigma$ credability intervals of the other runs.

\begin{equation}
    \label{eq:BIC}
    {\rm BIC} = k\ln(n) - 2\ln\left(\mathcal{L}\right)
\end{equation}

\begin{equation}
    \label{eq:AIC}
    {\rm AIC} = 2k - 2\ln\left(\mathcal{L}\right)
\end{equation}

We also did a seventh run which consists of the most probable configuration but with the eccentricity fixed to zero.  This run had a resulting log posterior probability of 392697.59, significantly lower than the 392716.16 from Run 2. It is well known that planets in such close orbits tend to show low eccentricity orbits, or even circular orbits, but since this signal is buried deep within the noise, it is not possible to fully constrain the orbital eccentricity, resulting in an eccentric orbit.

As an additional check we ran the same configurations including the Minerva data and found that in all three runs, while the signal is still recovered, significant noise is added to the fit, resulting in lower posterior probabilities. Thus we choose to exclude the Minerva RVs from the final analysis. Table \ref{tab:stats} show the aforementioned indicators for each run.

\begin{table}\label{tab:stats}
\caption{Model statistics for each run, compared against the baseline run. The row in bold face indicates the adopted run. The baseline run is a  linear fit to the data.}
\centering
\begin{tabular}{cccc}
\hline
run & Posterior & BIC & AIC \\\hline
Run 1 & 25.89 & 100.38 & 9.79 \\
{\bf Run 2} & {\bf 35.48} & {\bf -13.86} & {\bf -22.92} \\
Run 3 & 27.45 & 128.78 & 11.01 \\
Run 4 & 28.25 & 32.93 & -3.30 \\
Run 5 & 25.35 & -197.23 & 219.49 \\
Run 6 & 23.45 & 38.21 & 1.98 \\
baseline & 0 & 0 & 0 \\
\hline
\end{tabular}
\end{table}

\subsubsection{HD 18599 b}
\label{sec:planetb}

\begin{figure}
	\includegraphics[width=\columnwidth]{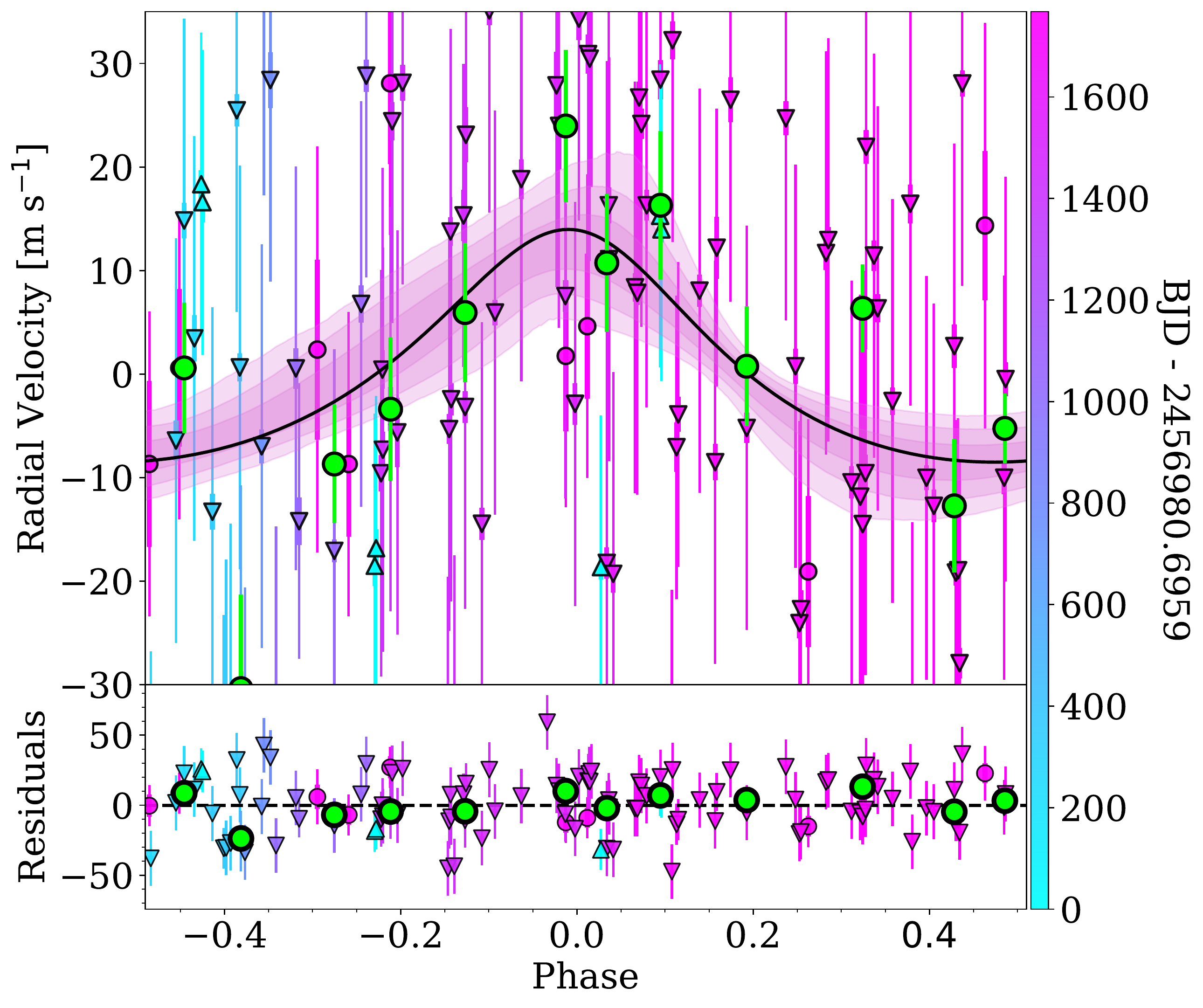}
    \caption{HD 18599 RVs folded at 4.137437 days. The black line represents the best fitting model while the purple shaded regions show the 1, 2 and 3$\sigma$ confidence regions of the model. The lower panel shows the residuals of the fit. Circles are HARPS\_pre data, upside down triangles show HARPS\_post data and upright triangles show the FEROS data. Green points show the data binned to 10 points in phase space. The colorbar encodes the observing time of each observation.}
    \label{fig:rvph}
\end{figure}

The modeling results show a radius ratio of $R_\text{p}/R_*=0.0311\pm0.0008$, resulting in a radius of 2.70$\pm$0.05~\re, making HD 18599 b a sub-Neptune planet, while the semi-amplitude, $K$, of 11$\pm$3~\ms, resulting in a derived mass of M$_\text{p}=25.5\pm4.6$~\me, meaning HD 18599 b has a density of $\rho_\text{p}=7.1\pm1.4$~\gccc.  Such a density therefore, is consistent with the planet maintaining an atmosphere that has a 24\% H$_2$O composition (see Section \ref{sec:discussion}). We summarize the model parameters in Table \ref{tab:planet} and show the phase folded RVs with the best fitting model and 1-3$\sigma$ credible intervals in Figure \ref{fig:rvph}.

We searched for additional signals in the RV residuals, but nothing significant was found. More sophisticated activity models are needed to further disentangle Keplerian signals from activity induced ones in noisy RV data.

\begin{figure}
	\includegraphics[width=\columnwidth]{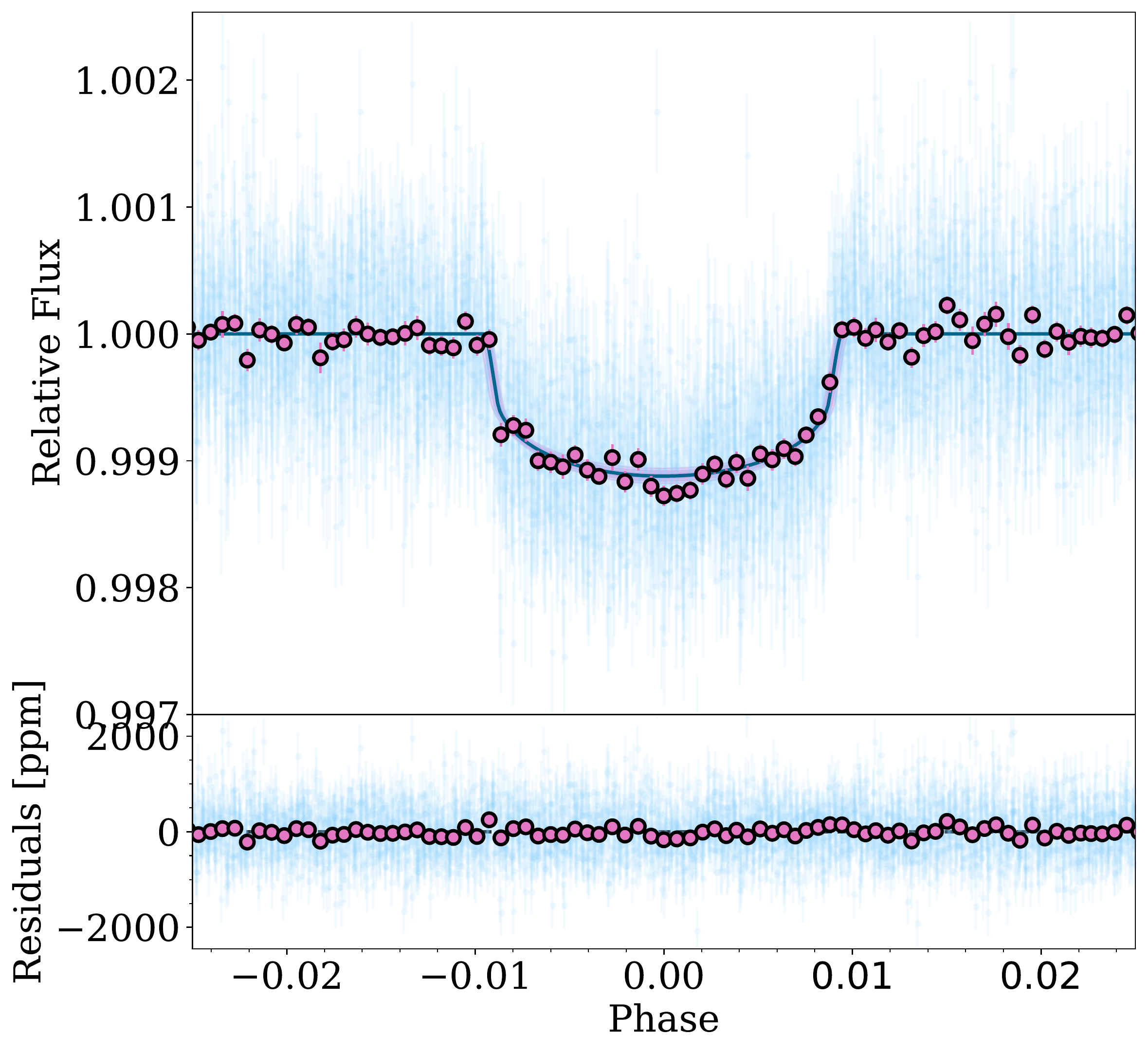}
    \caption{\tess{} light curve folded at 4.137437 days in light blue points with pink points showing the light curved binned to 50 points in phase space. The dark blue line represents the best fitting model and the pink shaded regions show the 1, 2 and 3$\sigma$ confidence regions of the model. The lower panel shows the residuals of the fit in ppm.}
    \label{fig:pmph}
\end{figure}

\subsubsection{TTV search}

The Kepler mission was responsible for the detection of a few thousand exoplanets, from which a handful of multi-planet systems were validated through the transit timing variation (TTV) technique \citep{holczer2016transit}.  Since the majority of Kepler stars were too faint for precise RV measurements, TTVs became key to derive planetary masses in many cases, thus validating the planetary systems \citep{lithwick2012extracting}. 
Moreover, this technique supported the idea that hot Jupiters are not part of multi-planet systems \citep{steffen2012kepler}, therefore setting major constraints on planetary migration models of giant planets.
The TTV method \citep{agol2005detecting} relies on the comparison between the measured and expected central transit times ($T_c$) from a linear model given by $T_n = T_0 + n \times P$, where $T_n$, $P$ and $n$ are the central transits, period, and transit number ($n=1,2,3,...$) respectively. Therefore, a deviation from the linear model may indicate the occurrence of dynamical interactions, where the most frequent cases are mean motion resonance planet-planet interaction and planet-star interaction leading to orbital tidal decays \citep{yee2019orbit}. 

\begin{figure}
	\includegraphics[width=\columnwidth]{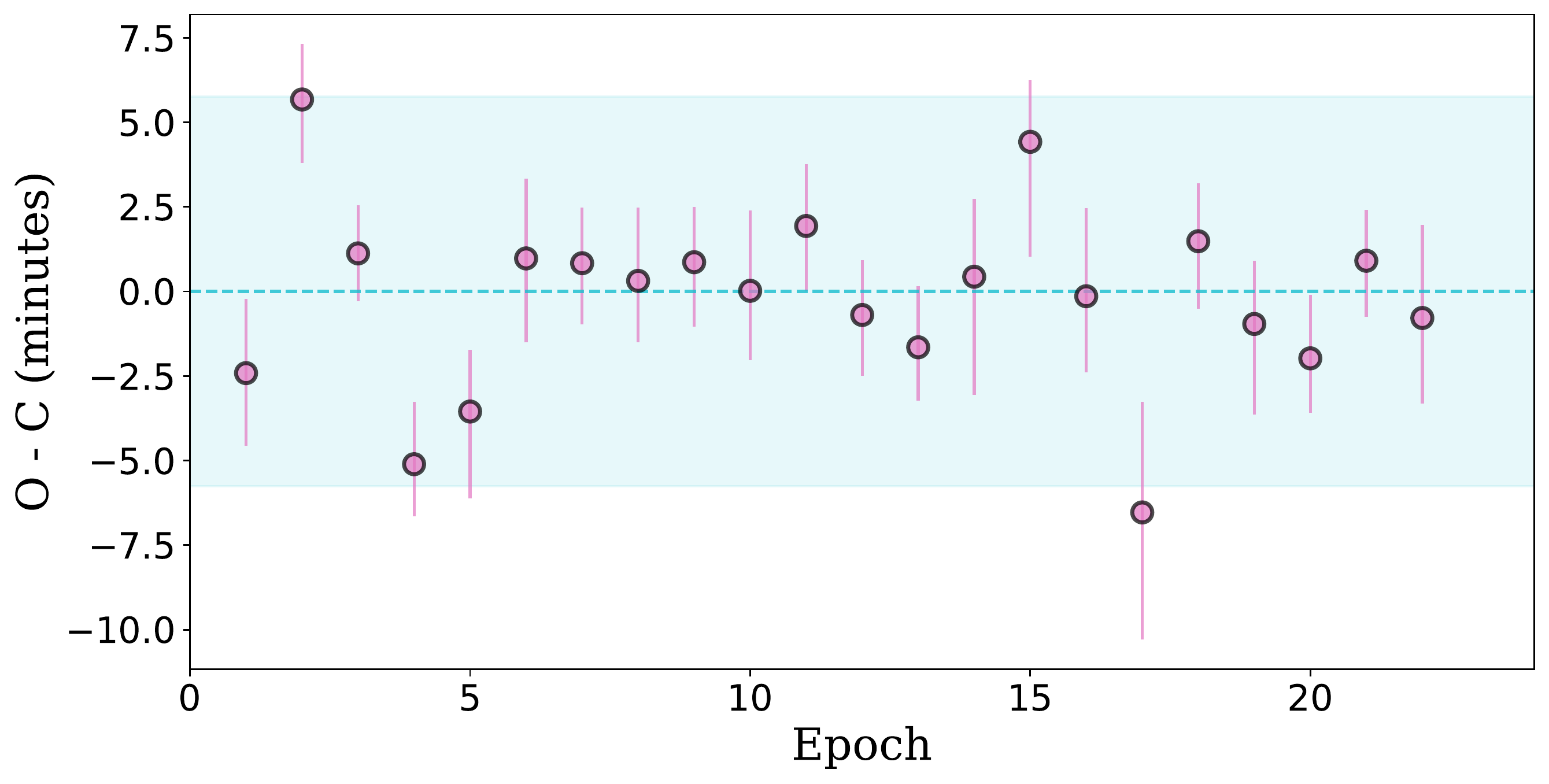}
    \caption{Observed minus computed mid-transit times for HD 18599b. The residuals (TTV) are shown considering the proposed ephemerides. The dashed cyan line shows zero variation and the shaded cyan region shows the 1$\sigma$ uncertainties on the linear ephemeris from \texttt{EMPEROR}. The plot shows no significant variation as all observations are within  1$\sigma$ of the best fitting ephemerides.}
    \label{fig:ttv}
\end{figure}

The TTV analysis was performed using \texttt{allesfitter}\footnote{https://www.allesfitter.com/} \citep{allesfitter} on the detrended light curve. Stellar and planetary parameters were fixed to the median posteriors from Tables \ref{tab:stellar} and \ref{tab:planet} respectively, except for $T_n$ which used $\mathcal{U}(T_n - 0.05, T_n + 0.05)$ for each $n$th transit. The O-C transit timing plot (Figure \ref{fig:ttv}) agrees with the linear model thus indicating no dynamical interaction with a second body. 

\begin{table}\label{tab:planet}
\caption{Fitted and Derived Parameters}
\centering
\begin{tabular}{ll}
\hline
Parameter                         & Prior                   \\ \hline
Orbital Parameters & \\
Fitted & \\ \hline
P [days]  & 4.137437$\pm$0.000004	\\\vspace{2pt}
K [ms$^{-1}$]         & 11$\pm$2	\\\vspace{2pt}
T$_c$ [JD]           & 2458726.9576$\pm$0.0004	\\\vspace{2pt}
$\sqrt{e}\sin(\omega)$ & -0.1$\pm$0.3 \\ 
$\sqrt{e}\cos(\omega)$ & 0.5$^{+0.1}_{-0.3}$ \\ 
$R_p/R_{*}$  &  0.0311$\pm$0.0008 \\\vspace{2pt}
b  & 0.58$\pm$0.11    \\\hline\vspace{2pt}
%%%
Derived & \\ \hline \vspace{2pt}
$\omega$ [rads]         & 0.2$^{+0.6}_{-0.2}$ \\\vspace{2pt}
e                 & 0.2$^{+0.1}_{-0.2}$  	\\\vspace{2pt}
$i$ [deg] & 87.7$^{+0.5}_{-0.7}$ \\\vspace{2pt}
\mpl [\me]& 25.5$\pm$4.6	\\\vspace{2pt}
\rpl [\re]& 2.70$\pm$0.05  \\\vspace{2pt}
$\rho_\text{p}$ [\gccc] & 7.1$\pm$1.4\\\vspace{2pt}
T$_p$ [JD]           & 2458726.41$^{+0.27}_{-0.23}$	\\\vspace{2pt}
$a/R_{*}$ & 13.78$^{+0.96}_{-1.08}$ \\\vspace{2pt}
a [AU] & 0.05$\pm$0.02 \\\vspace{2pt}
T$_\text{eq}$ [K] & 863$^{+21}_{-18}$ \\ 
Insolation [$S_\oplus$] & 145$^{+16}_{-12}$	\\ \hline
%%%
Stellar Parameters & \\ \hline
$\rho_{*}$ [\gccc]  &  2.87$^{+0.62}_{-0.66}$ \\\vspace{2pt}
$C_\text{FEROS}$  & 0.2$\pm$0.8 \\\vspace{2pt}
$C_\text{HARPS\_pre}$  & 1.0$^{+0.3}_{-0.8}$ \\\vspace{2pt}
$C_\text{HARPS\_post}$  & -2.1$^{+0.7}_{-0.0}$ \\\vspace{2pt}
$q_1$  & 0.68$\pm$0.17 \\
$q_2$  & 0.32$\pm$0.07 \\ \hline
% &&
RV Noise Parameters                  &                                    \\ \hline \vspace{2pt}
$\gamma_\text{FEROS}$ [\ms] & -86.3$^{+16.3}_{-14.5}$ \\\vspace{2pt}
$\sigma_\text{FEROS}$ [\ms] & 10.2$^{+5.0}_{-4.4}$  \\\vspace{2pt}
$\gamma_\text{HARPS\_pre}$ [\ms] & -2.7$^{+9.8}_{-9.6}$ \\\vspace{2pt}
$\sigma_\text{HARPS\_pre}$ [\ms] & 15.4$^{+4.5}_{-3.6}$ \\\vspace{2pt}
$\gamma_\text{HARPS\_post}$ [\ms] & -68.9$^{+11.2}_{-11.7}$ \\\vspace{2pt}
MA Coefficient $\omega$        & --       \\ 
MA Timescale $\beta$ [days]        & --       \\
AR Coefficient $\phi$        & --       \\ 
AR Timescale $\alpha$ [days]        & --       \\ \hline
Transit Noise Parameters & \\ \hline
offset [ppm]  & -0.000009$\pm$0.000003 \\
jitter [ppm]  & 291$\pm$4 \\
dilution  & 1 (fixed) \\ \hline
% &&
Acceleration Parameter            &       \\     \hline                     
$\dot{\gamma}$ [ms$^{-1}$/yr]   & 17$\pm$3    \\
\hline
\end{tabular}\\
The ARMA correlated noise model was not included in the selected model as described in the text.
\end{table}

%%%%%%%%%%%%%%%%%%%%%%%%%%%%%%%%%%%%%%%%%%%%%%%%%%%%%%%%%%%%%%%%%%%%%%%%%%%%%%

\section{Discussion}
\label{sec:discussion}

\begin{figure}
	\includegraphics[width=\columnwidth]{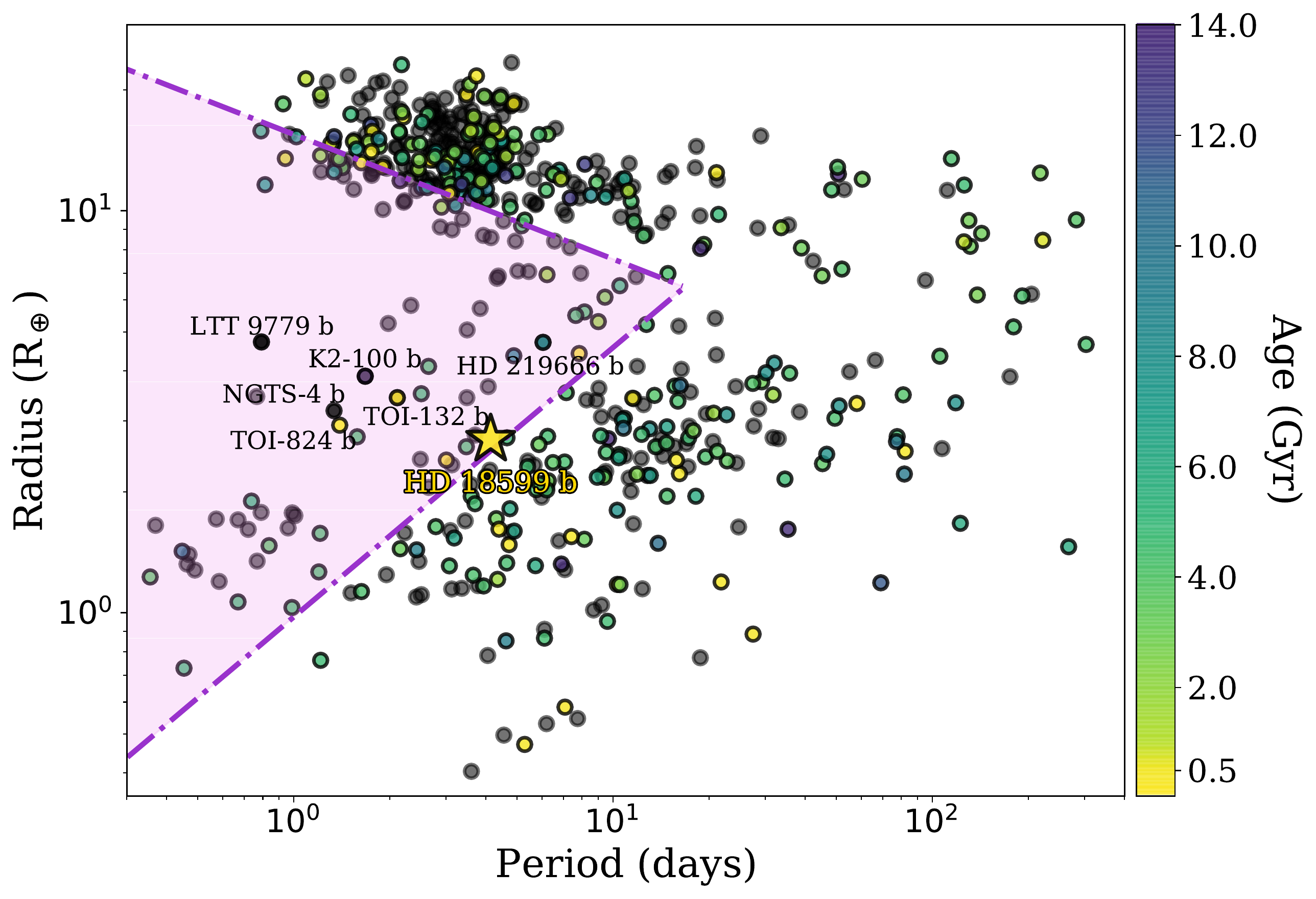}
    \caption{Period-radius diagram for planets with radii measured to a precision better than 5\%. Relevant discoveries are named, and HD 18599 is shown with a big yellow star at the edge of the desert. The pink shaded area shows the Neptune desert, with the edges defined by \citet{neptune-desert-edges}. The colours show the ages of the systems and black shows systems without an estimated age. We note that ages have not all been measured in an uniform way.}
    \label{fig:period-rad}
\end{figure}

We have confirmed the transit parameters found by de Leon et al. (2022, in prep.) and have characterized the orbit of the planet. Figure \ref{fig:period-rad} places HD 18599 b in the period-radius diagram, where we find it at the edge the Neptune desert. While the edge of the desert is fairly populated, the Figure shows that HD 18599 b is the youngest mini-Neptune discovered to date within this region, making this planet a key addition to the population.

\subsection{Possible Compositions}

\begin{figure}
	\includegraphics[width=\columnwidth]{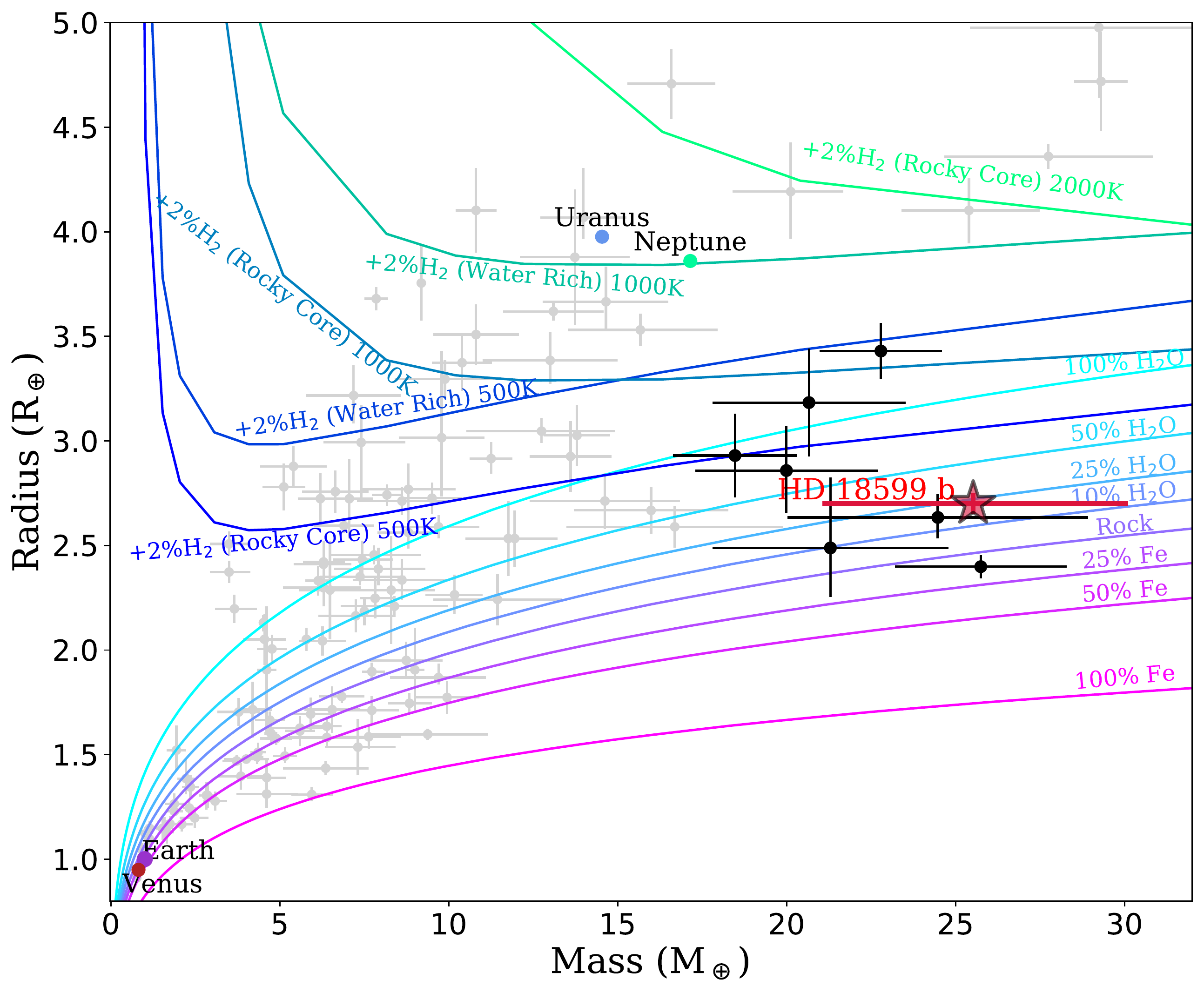}
    \caption{Mass-radius diagram for planets with masses and radii measured with a precision better than 20\% (gray circles) in the range R$_{\rm p}<5\,$R$_\oplus$ and M$_{\rm p}<30\,$M$_\oplus$, retrieved from the transiting exoplanets catalog \texttt{TEPCat} \citep{tepcat}. Black points show relevant discoveries. HD 18599 b is shown with a red star. Solid, colored lines show different bi-layer composition models from \citet{composition-models}, ranging from 100\% iron core planet to 100\% H$_2$O planet. Also five two-layer models from \citet{growth-models} are shown for 2\% H$_2$ envelopes at different temperatures and core compositions.}
    \label{fig:mass-rad}
\end{figure}

We measured the density of HD 18599 b to be 7.1$\pm$1.4 \gccc, and therefore we can compare this value to other known planets with similar masses and radii by placing it on a mass-radius diagram (Figure \ref{fig:mass-rad}).  HD 18599 b is shown by the red star, and it is placed against several two-layer composition models from \citet{composition-models} and five two-layer envelope models from \citet{growth-models}. Its position in the mass-radius diagram suggests the planet can be composed of 23\% H$_2$O, and 77\% rock (MgSiO$_3$) and iron core, while the 1$\sigma$ uncertainties allow the composition to vary from 10 to 41\% H$_2$O.

Due to the inherent degeneracies in three-layer bulk composition models with H/He envelopes on the sub-Neptune regime \citep[$2\leq$~R~$\leq4$~R$_\oplus$;][]{LopezFortney14}, we chose to further investigate the internal composition of HD 18599 b using the public tool \texttt{smint}\footnote{https://github.com/cpiaulet/smint} (Structure Model INTerpolator) and its interpolation and envelope mass fraction fitting package, first introduced by \citet{smint}. This code uses models from \citet{LopezFortney14}, \citet{composition-models}, and \citet{Aguichine21}, along with \texttt{emcee} to perform a MCMC fit of the H$_2$O or H/He mass fraction based on the planetary mass, radius, age and insolation flux. We set up the MCMC with 100 walkers and 10000 steps for all of the analyses. First we studied the H$_2$O mass fraction (WMF) using the \citet{Aguichine21} irradiated ocean worlds mass-radius relationships, where we found a WMF of 0.3 $\pm$ 0.1. The iron core mass fraction remains widely unconstrained with a value of 0.4 $\pm$ 0.3. The gas-to-core mass ratio analysis show that HD 18599 b has a 1$^{+14}_{-1}$\% mass fraction in H/He. The lack of an H/He envelope and the amount of water in the bulk composition argues for a steam atmosphere surrounding the planet.

The high mass regime in which HD 18599 b is found is poorly sampled, making the planet an ideal window into the physical and dynamical processes behind the evolution and formation of young Neptunes and their interactions with active stars.  Further work to better characterise the atmosphere of the planet are warranted, which would allow a better understanding of the internal composition of the planet and the physics of young planetary atmospheres.  Additional precision RV campaigns would also allow the possibility to discover additional companions in the system, in particular since it is known that small planet systems commonly harbor more than one planet (for example \citealt{2021arXiv211102416V, 2021AJ....162..215S}).

\subsection{HD 18599 as a Stellar Activity Model Testbed}

We tested different correlated noise models in order to account for the stellar activity signals appearing in the RV measurements.  These models included AR, MA, linear activity correlations, and a combination of all of these. We found that in this case, linear activity correlations gave rise to a higher posterior probability model, while ARMA models fell short of explaining the correlated noise, likely mainly due to the low observational cadence, since these models perform best under high cadence conditions. HD 18599 is an excellent testbed to apply more sophisticated activity models, also further encouraging more RV follow up work to be pursued.

\section{Conclusions}
\label{sec:conclusion}

Using \tess{} photometry and RVs from HARPS and FEROS we were able to characterize a dense planet orbiting the young star HD 18599.  The world is found to have an orbital period of 4.137437$\pm$0.000004 days, radius of 2.70$\pm$0.05 \re, (making it a mini-Neptune planet in size), and unusually high mass of 25.5$\pm$4.6 \me resulting in a planet with a density of 7.1$\pm$1.4 \gccc. The mass-radius diagram and bulk composition models suggests an ice rich core with a range of 10-41\% H$_2$O contribution, and a 1$^{+13}_{-1}$\% H/He envelope mass fraction. Further analysis shows a WMF of 33 $\pm$ 10\% and an unconstrained iron core mass fraction ranging from 10 to 70\%. HD 18599 b is the first mini-Neptune to orbit such a young (< 300 Myr) and bright (V$\sim$8.9) star, making it an ideal system to dive deep into evolutionary studies and atmospheric characterization.

\section*{Acknowledgements}

The authors became aware of a parallel effort on the characterization of TOI-179 by Desidera et al. in the late stages of the manuscript preparations. Only submissions to arxiv were coordinated, and no analyses or results were shared prior to the acceptance of the papers.

MINERVA-Australis is supported by Australian Research Council LIEF Grant 
LE160100001, Discovery Grants DP180100972 and DP220100365, Mount Cuba Astronomical 
Foundation, and institutional partners University of Southern 
Queensland, UNSW Sydney, MIT, Nanjing University, George Mason 
University, University of Louisville, University of California 
Riverside, University of Florida, and The University of Texas at Austin.

We respectfully acknowledge the traditional custodians of all lands 
throughout Australia, and recognise their continued cultural and 
spiritual connection to the land, waterways, cosmos, and community. We 
pay our deepest respects to all Elders, ancestors and descendants of the 
Giabal, Jarowair, and Kambuwal nations, upon whose lands the 
Minerva-Australis facility at Mt Kent is situated.

JIV acknowledges support of CONICYT-PFCHA/Doctorado Nacional-21191829.
JSJ greatfully acknowledges support by FONDECYT grant 1201371 and from the ANID BASAL projects ACE210002 and FB210003.
HRAJ acknowledges support from STFC grant ST/T007311/1
This research has made use of the NASA Exoplanet Archive, which is operated by the California Institute of Technology, under contract with the National Aeronautics and Space Administration under the Exoplanet Exploration Program.

\section*{Data Availability}
The data underlying this article are available in the article and in its online supplementary material.

%%%%%%%%%%%%%%%%%%%%%%%%%%%%%%%%%%%%%%%%%%%%%%%%%%



	\bibliographystyle{mnras}
\bibliography{paper} % if your bibtex file is called example.bib

%\begin{thebibliography}{99}
%\end{thebibliography}

%%%%%%%%%%%%%%%%%%%%%%%%%%%%%%%%%%%%%%%%%%%%%%%%%%

%%%%%%%%%%%%%%%%% APPENDICES %%%%%%%%%%%%%%%%%%%%%

%\appendix


%If you want to present additional material which would interrupt the flow of the main paper,it can be placed in an Appendix which appears after the list of references.

%%%%%%%%%%%%%%%%%%%%%%%%%%%%%%%%%%%%%%%%%%%%%%%%%%

% Don't change these lines
\bsp	% typesetting comment
\label{lastpage}
\end{document}